\documentclass[dvipsnames]{article}

\usepackage[utf8]{inputenc}
\usepackage{amsmath}
\usepackage{amsthm}
\usepackage{amssymb}
\usepackage{graphicx}
\usepackage{physics}
\usepackage{pgf,tikz,pgfplots}
\usepackage{orcidlink}
\usepackage[margin=1.1in]{geometry}
\usepackage{authblk}
\pgfplotsset{compat=1.16}
\usepackage{tabularx}
\usepackage{siunitx}
\usepackage{cleveref}

\usepackage{caption}
\usepackage{subcaption}
\usepackage[colorinlistoftodos]{todonotes}
\usepackage{xcolor}

\usepackage[normalem]{ulem}
\usepackage[acronym]{glossaries}

\usepackage[
backend=biber,
style=numeric-comp,
sorting=none
]{biblatex}

\newacronym{smf}{SMF}{single-mode fiber}
\newacronym{pdf}{PDF}{probability density function}
\newacronym{ao}{AO}{adaptive optics}

\title{Free-space model for a balloon-based quantum network}
\author[1,$\dagger$]{Ilektra Karakosta-Amarantidou}
\author[2,$\dagger$]{Raja Yehia}
\author[3]{Matteo Schiavon}
\affil[1]{Dipartimento di Ingegneria dell'Informazione, Universit\`a degli Studi di Padova, via Gradenigo 6B, IT-35131 Padova, Italy}
\affil[2]{ICFO - Institut de Ciencies Fotoniques, The Barcelona Institute of Science and Technology, Castelldefels, Spain}
\affil[3]{Sorbonne Universit\'e, CNRS, LIP6, F-75005 Paris, France}
\date{}

\begin{document}

\maketitle

\begin{abstract}
    Long-distance communication is one of the main bottlenecks in the development of quantum communication networks. Free-space communication is a way to circumvent exponential fiber loss and to allow longer communication distances. Satellite nodes are the main devices currently studied for free-space communication, but they come with downsides such as high cost and low availability. In this work, we study an alternative to satellites, namely aerial platforms such as high-altitude balloons. We provide a loss model to simulate the channel efficiency of balloon-to-ground, ground-to-balloon, and balloon-to-balloon communication channels, considering a large set of hardware parameters. We perform a parameter exploration to exhibit important trade-offs in these channels, as well as simulations of different quantum key distribution network architectures including balloon nodes. We demonstrate that balloons are a realistic alternative to satellites for free-space communications in national network architectures.
\end{abstract}
{\let\thefootnote\relax\footnotetext{$\dagger$ These authors contributed equally. For all inquiries, please contact: ikarakosta@proton.me}}\par
\section{Introduction}

Quantum networks hold the promise of improving the performance of current classical networks by increasing their security~\cite{SecurityQKD,Leaderelection,SecretSharing} or communication efficiency~\cite{RefCommComplexity}. They enable new functionalities such as, for instance, secure delegated computing~\cite{delegated1,DelegatedQC}, electronic voting~\cite{fedeVoting} and anonymous transmission~\cite{Anonymity}, while opening the way to distributed quantum computing and sensing~\cite{DANOS200773,shettell2022private}. Quantum networking has recently entered a rapid development phase with several experimental realizations of quantum networks~\cite{BristolQCity,ChinaQKDNetwork} as well as theoretical works on how to build and scale them up towards the so-called quantum internet~\cite{QIavision,IETFlinklayer}. \\

Long-distance communication remains a significant challenge in the establishment of a global-scale quantum network. As photons traverse fiber optic cables, losses become exponentially detrimental, rendering practical applications unfeasible for distances beyond a few tens of kilometers. Established constraints~\cite{PLOB,Takeoka_2014} provide fundamental thresholds for quantum communication through fiber optic. Quantum repeaters~\cite{Repeater1,Repeater2, sangouard2011quantum, azuma2022quantum}, akin to classical network signal amplifiers, offer a potential remedy to this issue. Despite notable strides in experimentation in recent years~\cite{Bhaskar_2020, azuma2022quantum,PompiliScience2021,Lago_Rivera_2021}, these repeaters have yet to attain the required technological maturity~\cite{repeaterNV,Ruf_2021,avis2022requirements}.

Over the past decade, much attention has focused on the exploration of free-space communication, particularly via satellite, as a complementary solution to circumvent these limitations. In free-space communication, loss primarily stems from diffraction, which increases quadratically with distance, in stark contrast to the exponential loss experienced in glass fiber media. Feasibility studies~\cite{Bonato2009,Bourgoin2013}, alongside ground-breaking experimental implementations of ground-to-ground~\cite{SchmittManderbach2007} and airplane-to-ground~\cite{Nauerth2013} free-space channels, have paved the way for exploratory endeavors in satellite-to-ground quantum communication. These efforts have included simulated quantum source experiments conducted aboard satellites~\cite{Vallone2015,Gnthner2017}. The launch of the Chinese satellite Micius~\cite{Lu2022} was a milestone that enabled the first-ever demonstrations of prepare-and-measure~\cite{Liao2017} and entanglement-based~\cite{Yin2020} quantum key distribution (QKD), as well as other quantum protocols like quantum teleportation~\cite{Ren2017}, utilizing a satellite terminal. 

Satellite communication, however, poses significant challenges, as they are costly to build, send to space and maintain. Their availability depends on weather and atmospheric conditions that are difficult to control. Pointing precision and fiber coupling through adaptive optics (AO) are active fields of research \cite{gruneisenAdaptiveOpticsEnabledQuantumCommunication2021,robertsPerformanceAdaptiveOptics2023} and remain very challenging. Moreover, 
low Earth orbit (LEO) satellites are preferred for quantum communications over geostationary Earth orbit (GEO) satellites because their closer proximity to Earth results in significantly lower transmission losses and reduced signal degradation. However, LEO satellites are visible from a ground station only for a few minutes per day. Although they can still be used for QKD networks, by developing strategies such as storing keys while the satellites are available and using them when needed, this does not fit the vision of a quantum Internet that could be accessed at any time for any application.

Several solutions are investigated, such as building smaller, less costly devices~\cite{Cubesat1,Cubesat2} or satellite constellations~\cite{SatConstellation}. In this work, we explore another alternative within free-space communications, namely quantum networks based on high-altitude balloons. High-altitude balloons, typically filled with helium or hydrogen, can be released into the atmosphere at an altitude of 18 to 38 km\footnote{\url{https://en.wikipedia.org/wiki/High-altitude_balloon}}. They can carry radio-equipment and are already used for research purposes and weather forecasting. They alleviate the availability constraint, while not completely removing it as usual weather balloons need daily maintenance. Other platforms such as high-altitude stations also known as atmospheric satellites, and hydrogen-powered aircrafts~\cite{AeroVironmentGO} have reported an endurance of a week. The construction of dedicated drone or balloon-based quantum nodes for quantum communication is still at a very early stage~\cite{Wang2013,Chu_2021,DUBEY2024100210,conradDronebasedQuantumCommunication2023,BookMajumdar2022} and only recently demonstrated experimentally for the first time \cite{tianExperimentalDemonstrationDroneBased2024}. In this article, we explore their feasibility by simulating a national quantum network where cities, equipped with fiber-based local networks, are linked through free-space balloon-based links. \\

We show that a balloon-based quantum network is indeed possible by simulating QKD rates between different users at distant locations. We build on the work started in~\cite{QCity}, in which we investigated the performance of a fiber-based metropolitan quantum network, called Quantum City, and continued in~\cite{yehia2023connecting},  where we simulated satellite communication between quantum cities. This was done by building a network simulator~\cite{github,githubMatteo} based on NetSquid~\cite{Netsquid,coopmans2021netsquid}. In this work, we provide a more accurate model of free-space quantum channels that we embed into the network simulator. We use this model to study and optimize different network properties, such as the placement of the balloons or the efficiency of balloon-to-balloon communication. In particular, we show that there are critical distances after which using a balloon link results in higher QKD rates than using a fiber link. These critical distances are around 80 km depending on the QKD protocol and the equipment used.

Our simulation tool, available on github~\cite{Newgithub}, allows us to place balloons at different locations and simulate the performance of several QKD protocols between users of a large network, while being flexible on many hardware parameters. Our free-space model for quantum communication is applicable to 1550 nm systems operating during both daytime and nighttime, while taking into account single-mode fiber (SMF) coupling, scintillation effects, adaptive optics correction, detector loss, tracking, and collection efficiency to model downlink and horizontal links. Our uplink model is based on a reciprocal modelling of the downlink channel. These models combine techniques from~\cite{scriminich2022optimal,robert2016impact,andrews2005laser,canuet2018statistical} and~\cite{vasylyev2018theory}.\\


\section{Free-space channel loss models} \label{sec:free_space_loss}

In this section, we briefly present our free-space loss models that we use to simulate balloon-based quantum communication networks. For readability, we leave most of the technical details to the appendices. \\

\begin{figure}[!ht]
    \centering
    \includegraphics[width=0.8\textwidth]{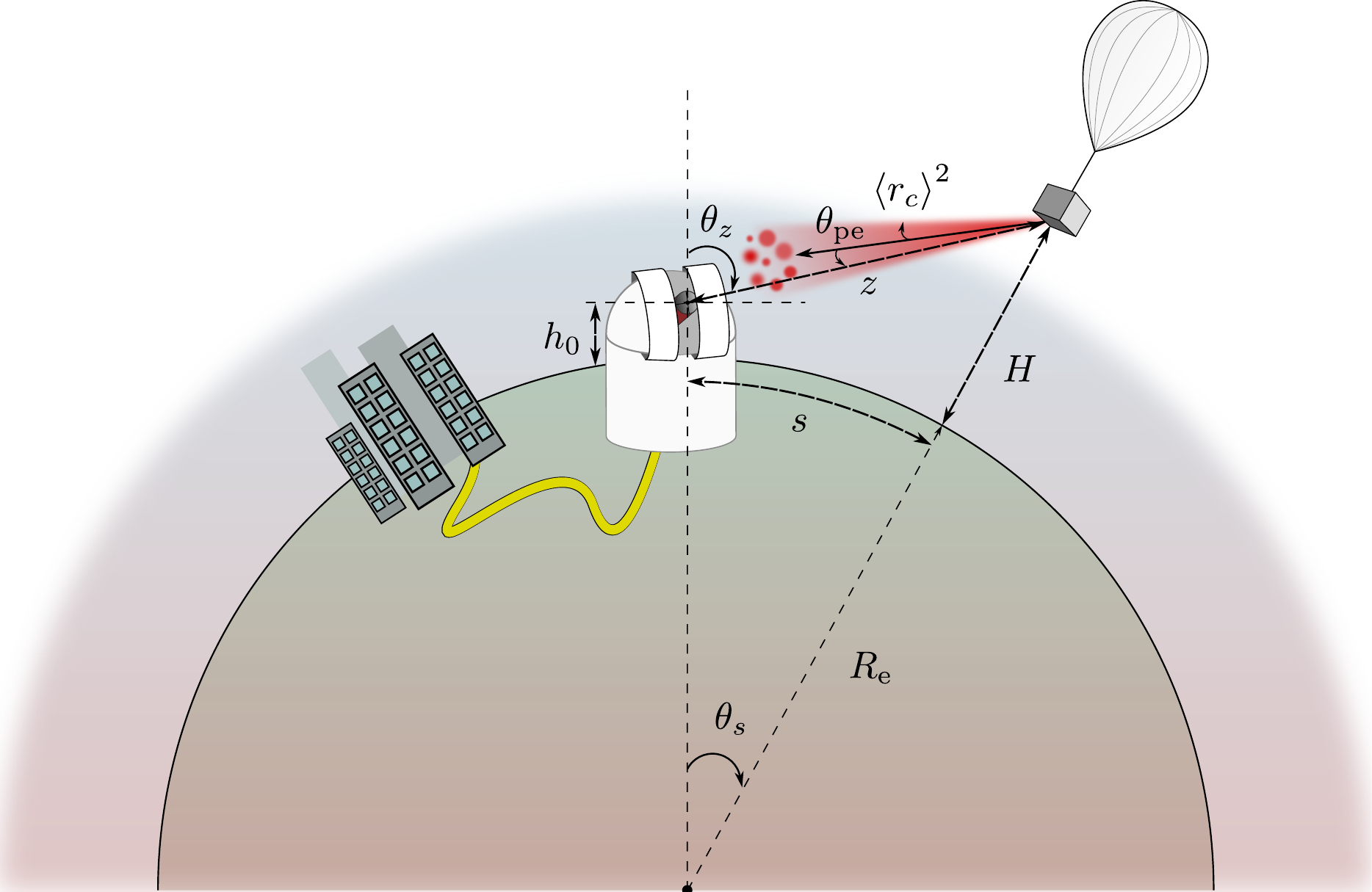}
    \caption{Example of network architecture simulated in this work: a free-space downlink of length $z$ between a balloon of altitude $H$ and a ground station of altitude $h_0$, with zenith angle $\theta_z$, pointing error $\theta_{\rm pe}$ and a beam wandering variance $\left\langle r_c^2\right\rangle$. $R_e$ is the radius of the Earth, $s$ the arc length between the ground station and the point on Earth below the balloon, and $\theta_s$ the corresponding subtending angle. The ground station is linked to a metropolitan fiber-based quantum network through an optical fiber.}   
    \label{fig:prezfig}
\end{figure}

For the different communication scenarios between ground stations and high-altitude balloons considered in this work, we define horizontal, downlink and uplink free-space loss models. To do so, we generalize the approach of the work presented in \cite{scriminich2022optimal}, which focused only on near-earth, ubran horizontal links. In particular, in the case of the downlink channel that is schematically represented in Fig.~\ref{fig:prezfig}, we introduce the dependence of the parameters quantifying the strength of atmospheric turbulence on the altitude of the high-altitude platform. Moreover, since we aim to simulate a variety of slanted atmospheric links including those at large zenith angles, we consider formulas that are valid for weak to strong atmospheric turbulence regimes. Then, we exploit the formulations of the downlink model to determine the losses in the uplink case through the principle of reciprocity \cite{shapiro2012reciprocity,puryear2013reciprocity,robert2016impact,lognonePhaseEstimationPointahead2023}. The considerations relative to the uplink channel are presented in Appendix~\ref{sec:uplink}. Finally, the horizontal channels examined in our simulations correspond to links between balloons placed at altitudes 18 km or higher, where the atmosphere is less dense. We thus use formulas valid in the weak turbulence regime in the horizontal case. 

With the above in mind, we assume that the total channel efficiency of each link is given by
\begin{equation}
    \eta_{\rm free-space} = \eta_{\rm atm} \cdot \eta_{\rm Rx}\, ,
\end{equation}
where $\eta_{\rm atm}$ is the atmospheric transmittance and $\eta_{\rm Rx}$ are the losses at the receiver defined as
\begin{equation}
    \eta_{\rm Rx} = \eta_{D_{\rm Rx}} \cdot \eta_{\rm SMF},
\end{equation}
with $\eta_{\rm D_{Rx}}$ the receiver collection efficiency and $\eta_{\rm SMF}$ the fiber-coupling efficiency. Both $\eta_{\rm D_{Rx}}$ and $\eta_{\rm SMF}$ depend on time and, therefore, we will use their \acrfull{pdf} to extract the loss probability of each photon traveling in the channel. In the following, we take a closer look on each one of these contributions to the overall channel losses.

\subsection{Atmospheric transmittance}

The atmospheric transmittance $\eta_{\rm atm}$ includes losses caused by effects such as molecular absorption and scattering, continuum absorption and aerosol extinction. In our analysis, we make use of the commonly adopted LOWTRAN software package \cite{kneizys1988users}, and in particular the Python integration of \cite{piccia_lowtran}, which is based on experimental measurements to predict atmospheric transmittance values over a wide spectral range, for different geographical areas or meteorological and aerosol conditions.

\begin{figure}[h!]
    \centering
    \includegraphics[width=0.6\textwidth]{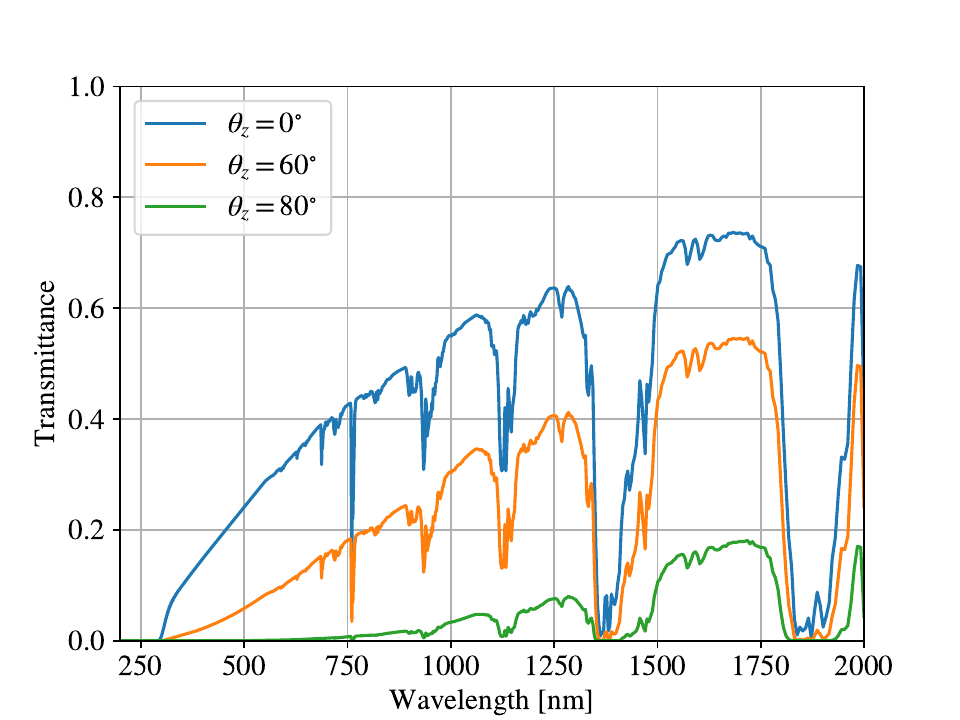}
    \caption{Amospheric transmittance generated by LOWTRAN for different zenith angles with ground station at sea level and aerial platform at 20 km.}
    \label{fig:transmittance_example}
\end{figure}

Fig.~\ref{fig:transmittance_example} demonstrates that working at 1550 nm is relevant for free-space communication, since, apart from compatibility with existing fiber infrastructures, the atmospheric transmittance is close to its maximum around this wavelength. As expected, the transmittance decreases when considering higher zenith angles $\theta_z$, as the thickness of the atmosphere increases when $\theta_z$ becomes larger (see Fig.~\ref{fig:prezfig}).

\subsection{Probability distribution of collection efficiency} \label{sec:coll_eff}

The collection efficiency $\eta_{D_{\rm Rx}}$ of a receiving telescope is its ability to collect light in the presence of turbulence-induced scintillation, beam broadening and beam wandering, as well as mechanical pointing error between the transmitter and receiver. Here, we give a high-level description of how these effects influence the collection efficiency. More details can be found in Appendix~\ref{sec:app_pdt}. \\

Atmospheric turbulence effects are caused by random fluctuations of the refractive index of the atmosphere which in turn perturb the wavefront of propagating beams causing coherence loss. There exist various models which describe the power spectral density of the refractive index fluctuations over the spatial frequency $\kappa$, with the most famous one being the Kolmogorov spectrum \cite{andrews2005laser}
\begin{equation}
\Phi_{\mathrm{n}}(\kappa, h)=0.033 C_{\mathrm{n}}^2(h) \kappa^{-11 / 3},
\end{equation}
where $C_n^2(h)$ is the refractive index structure constant depending on the altitude from sea-level $h$. A widely used model for $C_n^2(h)$ is the Hufnagel-Valley (HV) model \cite{andrews2005laser}
\begin{equation}
    C_n^2(h) = 0.00594\left(\frac{\bar{v}_{\rm wind}}{27}\right)^2\left(10^{-5} h\right)^{10} \exp \left(-\frac{h}{1000}\right)+2.7\cdot 10^{-16} \exp \left(-\frac{h}{1500}\right)+ C_n^2(0) \exp \left(-\frac{h}{100}\right),
\end{equation}
where $\bar{v}_{\rm wind}$ is the average transverse wind velocity and $C_n^2(0)$ the reference parameter of $C_n^2(h)$ at ground level.

Scintillation refers to the fluctuation of the intensity (irradiance) of a beam due to the turbulent channel. The turbulence-induced beam broadening is additional to the one caused by diffraction and impacts the long-term size of a propagating Gaussian beam $W_{\rm LT}(z)$. This effect can be thought of as the combination of small-scale contributions, leading to an increased size of the short-term radius $W_{\rm ST}(z)$, and a large-scale beam wandering, characterized by a random displacement of the beam's centroid over the aperture of the receiver. The variance of this random wandering of the beam $\left\langle r_c^2\right\rangle$ (see Fig.~\ref{fig:prezfig}), the short-term radius $W_{\rm ST}(z)$ and the turbulence-broadened beam $W_{\rm LT}(z)$ are linked by the relation
\begin{equation}
    W_{\rm LT}^2(z) = W_{\rm ST}^2(z) + \left\langle r_c^2\right\rangle.
\end{equation}

The pointing error due to misalignment of the communicating terminals is additional to the turbulence-induced beam wandering. If we assume an angular pointing error $\theta_{\rm pe}$, the wandering variance should be substituted by 
\begin{equation}\label{eq:wander_no_track}
    \sigma_{\rm wander}^2 = (z \theta_{\rm pe})^2 + \left\langle r_c^2\right\rangle.
\end{equation}
In the same vein, in case an active tracking mechanism is present at the receiving telescope, it can be incorporated in Eq.~\eqref{eq:wander_no_track} as 
\begin{equation}
    \sigma_{\rm wander}^2 = \left[\left(z \theta_{\rm pe}\right)^2 + \left\langle r_c^2\right\rangle\right]\cdot(1 - \eta_{\rm tr}),
\end{equation}
where $\eta_{\rm tr}$ is the efficiency of the tracking mechanism. \\

 \begin{figure}[!ht]
    \centering
    \begin{subfigure}[b]{0.49\textwidth}
        \centering
        \includegraphics[width=\textwidth]{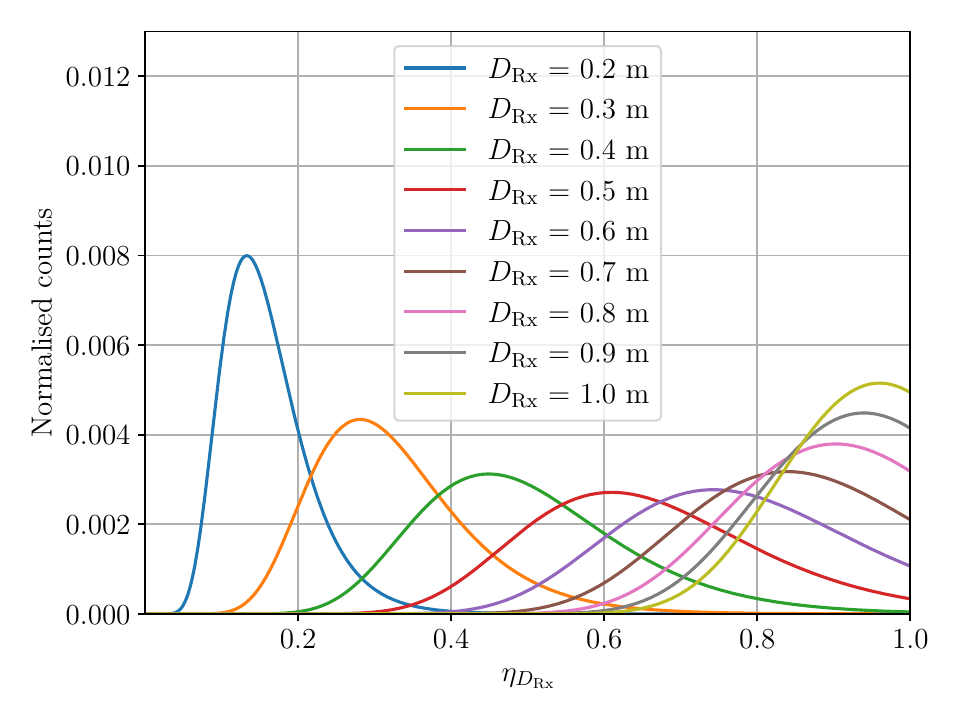}
        \caption{Collection efficiency for different sizes of the receiver aperture. The tracking efficiency is fixed to $\eta_{\rm tr} = 85\%$.}
        \label{fig:coll_eff_pdf_drx}
    \end{subfigure}    
    \hfill
    \begin{subfigure}[b]{0.49\textwidth}
        \centering
        \includegraphics[width=\textwidth]{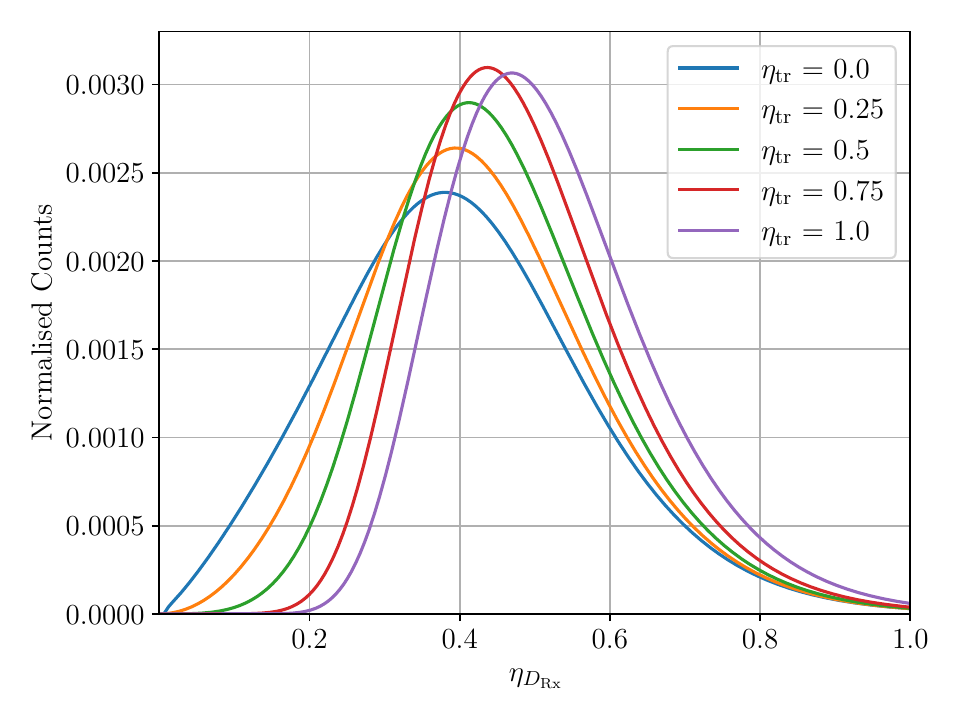}
        \caption{Collection efficiency for different tracking efficiencies. The receiver aperture is fixed to $D_{\rm Rx} = 0.4$.}
        \label{fig:coll_eff_pdf_track}
    \end{subfigure}
     \caption{Receiver collection efficiency PDF in a downlink channel with ground station at $h_0 = 20$ m, aerial platform at $H = 20$ km, zenith angle $\theta_z = 70^{\circ}$, initial beam waist radius $W_0 = 0.1$ m and pointing error $\theta_{\rm pe} = 2 \ \mu$rad. We notice that the average collection efficiency increases while the receiver aperture diameter becomes larger or when a tracking beam tracking mechanism with higher efficiency is employed.}
     \label{fig:collection_eff}
\end{figure}

According to \cite{vasylyev2018theory}, the collection efficiency \acrshort{pdf} for short distances or in the weak irradiance fluctuations regime is mainly impacted by beam wandering, and has the form of a log-negative Weibull distribution. On the other hand, for longer distances and strong scintillation conditions, beam-spot distortion is more profound and the PDF resembles more closely the truncated log-normal distribution. A general relation that is valid from weak to strong scintillation regimes can be found using the law of total probability through 
\begin{equation} \label{eq:pdt_vec}
    P_{D_{\rm Rx}}(\eta_{\rm D_{Rx}})=\int_{\mathbb{R}^2} P\left(\eta_{\rm D_{Rx}} \mid \boldsymbol{r}\right) \rho\left(\boldsymbol{r}\right) d^2 \boldsymbol{r},
\end{equation}
with $\boldsymbol{r}$ the displacement of the beam from the receiver aperture center described by a normal distribution $\rho(\boldsymbol{r})$ of average $0$ and variance $\sigma_{\rm wander}^2$,
and $P(\eta_{D_{\rm Rx}}|\boldsymbol{r})$ a distribution which has been demonstrated to approximate the truncated log-normal distribution. The final form of Eq.~\eqref{eq:pdt_vec} and its full derivation are given in Appendix~\ref{sec:app_pdt_deriv}.

This PDF effectively varies between the log-negative Weibull and the truncated log-normal distribution, depending on the strength of the turbulence. For the horizontal links simulated in this work, we only consider the effects of beam wandering since at these altitudes the strength of the atmospheric turbulence is quite small. In Fig.~\ref{fig:collection_eff}, we see an example of the form of the collection efficiency PDF for different receiver aperture diameters $D_{\rm Rx}$ and tracking efficiencies $\eta_{\rm tr}$ in a downlink channel.

\subsection{Probability distribution of fiber coupling efficiency} \label{sec:smf_pdf}

In this section, we describe the model used for the coupling efficiency into an SMF. The model, presented in greater detail in Appendix~\ref{sec:app_smf_pdf}, includes the statistical properties of the coupling efficiency and is related to the case of a downlink channel, where the receiving ground station uses an AO system for the correction of the wavefront aberrations in the incoming beam. For horizontal channels, where the studied altitudes correspond to weak turbulence conditions, we assume that the receiving balloons do not have AO and we only consider the average value of the instantaneous coupling efficiency. \\

The coupling efficiency $\eta_{\rm SMF}$ is given by the product of three terms: 
\begin{equation} \label{eq:eta_smf2}
    \eta_{\rm SMF} = \eta_0 \eta_{\rm \chi} \eta_{\phi},
\end{equation}
where $\eta_0$ is the maximum achievable coupling efficiency when there is no turbulence, which depends on the obstruction ratio $\alpha_{\rm obs}= D_{\rm obs}/D_{\rm Rx}$, where $D_{\rm obs}$ is the diameter of the central obstruction of the receiving telescope, $\eta_{\rm \chi}$ corresponds to the contribution of scintillation, and $\eta_{\phi}$ is the coupling efficiency due to aberrations in the phase of the wavefront that can be partially corrected by AO. Of these three terms, only $\eta_{\phi}$ is statistical in nature. The PDF of $\eta_{\rm SMF}$ can be extracted by breaking down the random phase $\phi$ into a set of orthonormal polynomials over the receiver aperture, as explained in Appendix~\ref{sec:phase_term}. Each of these polynomials, of azimuthal and radial order $n$ and $m$ respectively, corresponds to a different aberration of the phase whose strength is quantified by a coefficient $b_j$. The action of the AO system is to attenuate each aberration by an attenuation factor $\gamma_j$. The values of the attenuation factors depend on parameters such as the gain $K_I$, delay $\tau$, integration time $T$ and maximum order of correction $N_{\rm AO}$ of the control loop that is part of the AO system.

\begin{figure}[!ht]
    \centering
    \begin{subfigure}[b]{0.49\textwidth}
        \centering
        \includegraphics[width=\textwidth]{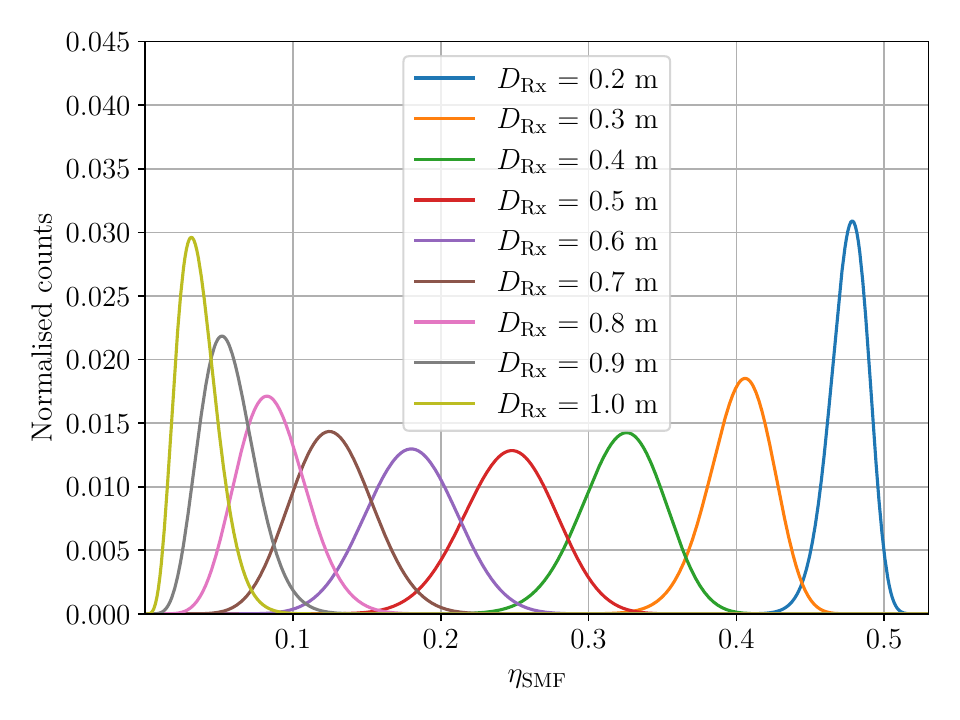}
        \caption{Coupling efficiency PDF for different sizes of the receiver aperture. The maximum corrected radial order by the AO system is fixed to $N_{\rm AO} = 6$.}
        \label{fig:smf_pdf_drx}
    \end{subfigure}    
    \hfill
    \begin{subfigure}[b]{0.49\textwidth}
        \centering
        \includegraphics[width=\textwidth]{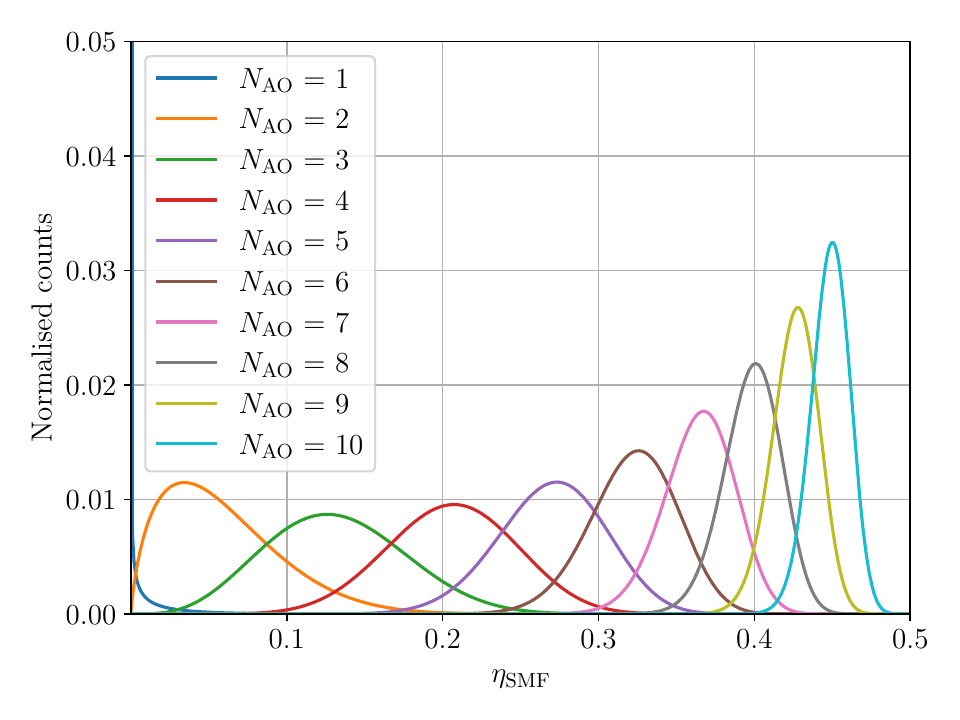}
        \caption{SMF coupling efficiency PDF for different maximum corrected radial order by the AO system. The receiver aperture is fixed to $D_{\rm Rx} = 0.4$.}
        \label{fig:smf_pdf_ao}
    \end{subfigure}
     \caption{Coupling efficiency PDF in a downlink channel with ground station at $h_0 = 20$ m, aerial platform at $H = 20$ km, zenith angle $\theta_z = 70^{\circ}$, initial beam waist radius $W_0 = 0.1$ m and obstruction ratio at the receiver $\alpha_{\rm obs} = 0.3$.}
     \label{fig:smf_pdf}
\end{figure}

According to the results of \cite{canuet2018statistical,scriminich2022optimal}, we define $\xi(t)$ as the instantaneous sum of the corrected coefficients by
\begin{equation}
    \xi(t) = \sum_{j} \gamma_j^2 \left\langle b_j^2 \right\rangle,
\end{equation}
where the amount of terms in the sum depends on the maximum radial order $N_{\rm max}$ considered in the wavefront decomposition. Quantity $\xi$ describes the residual error after correction of the wavefront from the AO system. The PDF of $\xi$ is 
\begin{equation} \label{eq:eta_xi}
P_{\xi}(\xi)=\frac{1}{\pi} \int_0^{\infty} \frac{\cos \left[\sum_{j} \frac{1}{2} \arctan \left(2 \gamma_j^2 \left\langle b_j^2 \right\rangle u\right)-\xi u\right]}{\prod_{j}\left[1+4 u^2\left(\gamma_j^2 \left\langle b_j^2 \right\rangle\right)^2\right]^{1 / 4}} \mathrm{~d} u
\end{equation}
and, based on the above, the PDF of $\eta_{\rm SMF}$ is
\begin{equation} \label{eq:smf_pdf}
P_{\mathrm{SMF}}\left(\eta_{\mathrm{SMF}} \mid \eta_{\mathrm{max}}\right)=\frac{1}{\eta_{\mathrm{SMF}}} P_{ \xi}\left[\log \left(\frac{\eta_{\max }}{\eta_{\mathrm{SMF}}}\right)\right],
\end{equation}
where $\eta_{\rm max} = \eta_0 \eta_{\chi}$. Examples of the coupling efficiency PDF for different receiver apertures and orders of AO correction in a downlink channel are shown in Fig.~\ref{fig:smf_pdf}. We notice that, contrary to the collection efficiency, the coupling efficiency decreases as the receiver aperture becomes larger (Fig.~\ref{fig:smf_pdf_drx}). This implies that for a specific $N_{\rm AO}$, there is an optimal $D_{\rm Rx}$ that minimizes the overall losses at the receiver. This type of trade-off is analyzed in detail on \cite{scriminich2022optimal}. Alternatively, a higher order of correction can be employed to mitigate the effects of wavefront distortions on fiber-coupling efficiency (Fig.~\ref{fig:smf_pdf_ao}) when a particular size of aperture has to be used.

\section{Simulation setup and parameters}
\label{sec:setup}

We will now apply the free-space loss models described in Sec.~\ref{sec:free_space_loss} to explore the possibility of using high-altitude balloons in real-life scenarios, following up on the work from \cite{QCity} and \cite{yehia2023connecting}. In this section, we recall the parameters of interest and describe our simulation tool.\\

\begin{figure}[h!]
    \centering
    \includegraphics[width=0.7\linewidth]{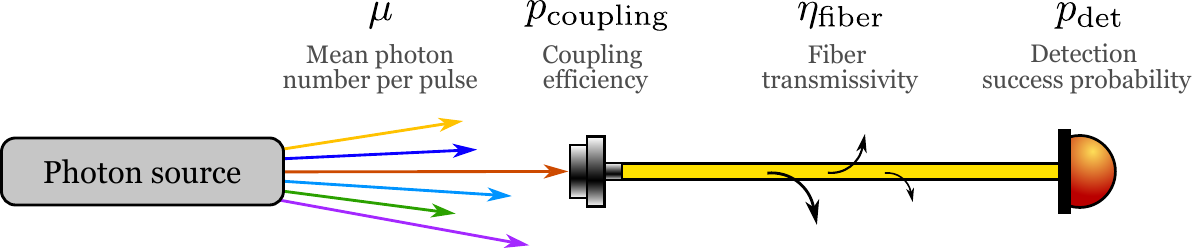}
    \caption{Loss model for the creation and reception of a photon through an optical fiber.}
    \label{fig:Photonlossmodel}
\end{figure}

In~\cite{QCity}, we introduced the concept of a quantum city, a star-like network with a powerful central node, called the Qonnector, and end users, called Qlients, with minimal hardware requirement. Namely, we only assume the Qlients to be able to create, manipulate and measure one photonic qubit at a time. This requirement is fundamental for large-scale adoption of quantum network architectures, as we cannot expect all end users to have access to powerful quantum devices. Qonnectors on the other hand, can manipulate several qubits at a time, perform two-qubit measurements, route qubits from one Qlient to another and, in the context of this work, act as ground stations for free-space communication. We characterized different configurations for Qonnectors and Qlients in a metropolitan network. To do so, we modeled imperfect quantum processes through erasure channels, depolarizing channels and dephasing channels. Effectively, for every quantum operation, there is a probability that the qubit is lost or flipped. We show in Fig.~\ref{fig:Photonlossmodel} the loss model used in this work, inspired from~\cite{QCity}, for creating, sending and receiving a photon through an optical fiber. Since we focus our study on the overall channel efficiency, we do not consider noise in the simulation of the channels, but rather fix it to a realistic value when relevant.

As we focus on photonic communication at wavelength $\lambda = 1550$ nm, we fix the fiber transmissivity at $\eta_\text{\rm fiber} = 0.18$ dB/km. We consider two different types of detectors: superconducting nanowire single-photon detectors (SNSPD) with efficiency $p_{\rm det}=0.85$ in the ground stations, and single-photon avalanche diodes (SPADs) with efficiency $p_{\rm det}=0.25$ in the aerial platforms. Indeed, SNSPDs, while having a higher efficiency, require cryogenics that cannot be easily embedded in an aerial platform. SPADs, on the other hand, are commercially available and fairly compact at the expense of a lower efficiency. 
In Table~\ref{tab:baselineparameters}, we show all the other parameters that are fixed for the rest of this work. We emphasize, however, that they can be freely changed in our software. In the next sections, we study the influence and interplay of specific parameters of the free-space channel, namely the channel length $z$, the zenith angle $\theta_z$, the beam waist radius at the transmitter $W_0$, the aperture of the receiving telescope $D_{\rm Rx}$ and the maximum radial order corrected by adaptive optics $N_{\rm AO}$. Geometrical considerations, useful to compute the different distances and heights of the balloons with respect to a spherical Earth, are given in Appendix~\ref{sec:geocons}. \\

\begin{table}[!ht]
    \centering
\begin{tabular}{l|c|l}
    Symbol & Value & Description  \\ \hline
    $\lambda$ & $1550$ nm & Wavelength\\
    $\eta_\text{fiber}$ & $0.18$ dB/km & Fiber loss per kilometer \\ 
    $p_{\rm det}^{\rm SNSPD}$ & $0.85$  & Detector efficiency at the ground  \\
    $p_{\rm det}^{\rm SPAD}$ & $0.25$  & Detector efficiency in the aerial platform \\
    $h_0$ & $20$ m & Altitude of the ground station \\
    $C_n^2(0)$ & $ 9.6 \cdot 10^{-14}\ \mathrm{m}^{-2/3}$ & Refractive index structure constant at ground level \\
    $\eta_{\rm tr}$ & $80$\% & Tracking efficiency \\
    $\alpha_{\rm obs}$ & 0.3 & Obstruction ratio of the receiving telescope \\
    $\bar{v}_{\rm wind}$ & 10 m/s & Average transverse wind velocity \\
    $\theta_{\rm pe}$ & $1\mu$rad & Pointing error \\
    $K_I$ & 1 & Integral gain of the AO system \\
    $\tau$ & $2\cdot10^{-3}$ s& Control delay of the AO system \\
    $T$ & $1\cdot10^{-3}$ s & Integration time of the AO system \\
    $N_{\max}$ & 150 & Maximum radial order in wavefront decomposition\\
\end{tabular}
    \caption{Baseline simulation parameters}
    \label{tab:baselineparameters}
\end{table}

\begin{figure}[!ht]
    \centering
    \includegraphics[width=0.8\linewidth]{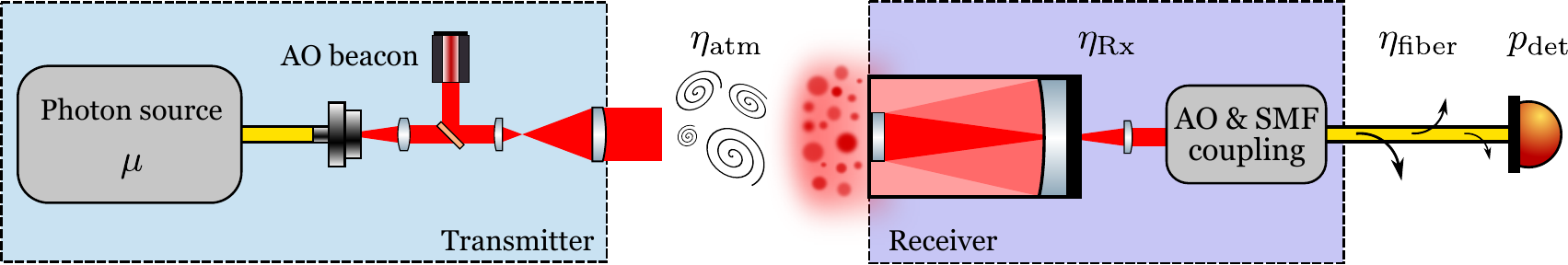}
    \caption{Loss model for the creation and reception of a photon through a combination of free-space and fiber channels.}
    \label{fig:Totallossmodel}
\end{figure}

We use a quantum network simulation tool, called NetSquid~\cite{Netsquid,coopmans2021netsquid}. Some modules, such as the creation of nodes performing basic quantum operations linked by quantum channels that can apply loss models, are included in NetSquid, combined with discrete event simulation, which is a modelling paradigm suited to network systems. This facilitates the simulation of realistic quantum networks. To learn more about the use of NetSquid, we encourage the reader to consult the NetSquid website~\cite{Netsquid}.  On top of this network simulator, we extend the open-source software of~\cite{QCity,yehia2023connecting}, in which the authors developed functions to create quantum cities and implemented some network protocols such as QKD. We improve on this by adding our free-space loss model and creating routines to define high-altitude balloon nodes. Compared to previous work, this model has a more precise characterization of the atmospheric turbulence and the coupling efficiency at the receiver under daylight conditions. This allows us to create a balloon-based national network and to estimate its performance on executing basic network protocols using realistic settings. The model that we use when free-space channels are introduced between nodes is illustrated in Fig.~\ref{fig:Totallossmodel}.
\\

The code used in our work is available on GitHub \cite{Newgithub} and can be used to test different configurations, with all the parameters described so far being tunable. Note however, that there are some restrictions on the parameter values for which our model produces accurate results, due to the assumptions that we considered during the formulation of the model. First of all, we consider the aperture averaging assumption, which is a reduction in the scintillation experienced by the receiver when its diameter is larger than a quantity called correlation width. It implies that a sufficiently big receiving aperture must be used (Appendix.~\ref{sec:collection_scint}). Then, we make the hypothesis that the residual wavefront error is low, which entails that the maximal order of correction of the AO system needs to be large enough in the downlink and uplink channels (Appendix.~\ref{sec:AvrgFiberCoupl}). Finally, we assume that the beam wandering variance is small enough, so that the beam is not exiting the receiver aperture, which effectively restricts the value of the pointing error (Appendix.~\ref{sec:beam_wandering}). The precise values corresponding to these restrictions depend on the parameters of the channel, and become limiting only for very long distances and high zenith angles. Our code raises warnings when this is the case.

\section{Results}

\subsection{Parameter exploration}
\begin{figure}[!ht]
 \label{fig:VerticAndHoriz}
     \centering
          \begin{subfigure}[b]{0.49\textwidth}
         \centering
         \includegraphics[width=0.6\textwidth]{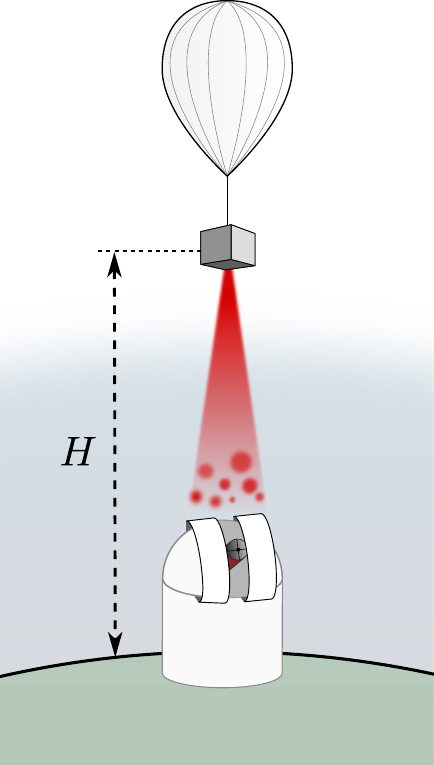}
         \caption{}
         \label{fig:vertic}
     \end{subfigure}    
     \hfill
\begin{subfigure}[b]{0.49\textwidth}
         \centering
         \includegraphics[width=\textwidth]{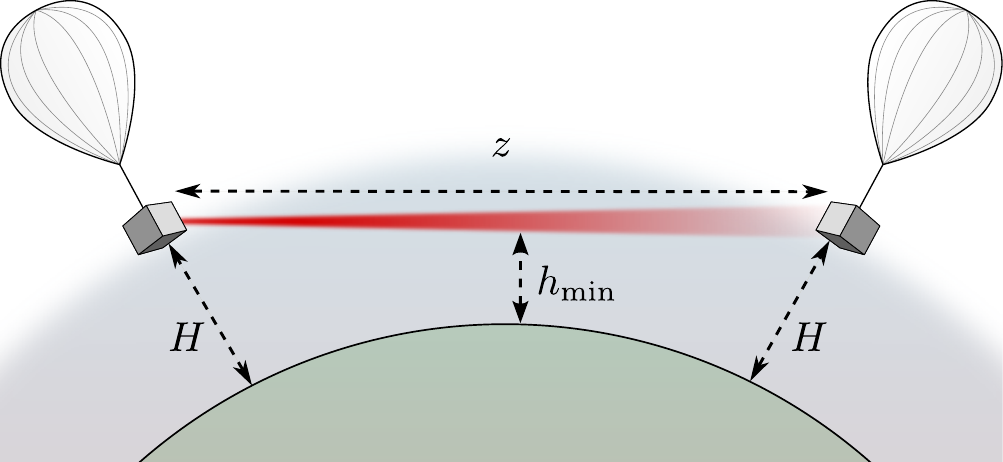}
         \caption{}
         \label{fig:horiz}
     \end{subfigure}
     \caption{Schematic representation of (a) a vertical downlink channel and (b) a horizontal channel.}
\end{figure}

Before analyzing a full-scale network simulation, we explore possible configurations in terms of number of balloons, their positions, the receiver and transmitter aperture sizes, as well as the maximum order of correction by the AO system employed in the ground station. For this step, we mostly focus on the mean channel efficiency, taking into account the considerations presented in the previous sections. We assess the quality of the Netsquid embedding of the loss model by comparing the simulation results with the theoretical value of the free-space channel efficiency. When not specified, the channel parameters are given in Table~\ref{tab:baselineparameters}. We start by studying the vertical downlink (Fig.~\ref{fig:vertic}) and horizontal (Fig.~\ref{fig:horiz}) channels. \\

\begin{figure}[!ht]
 \begin{subfigure}{.5\textwidth}
    \centering
    \includegraphics[width=\textwidth]{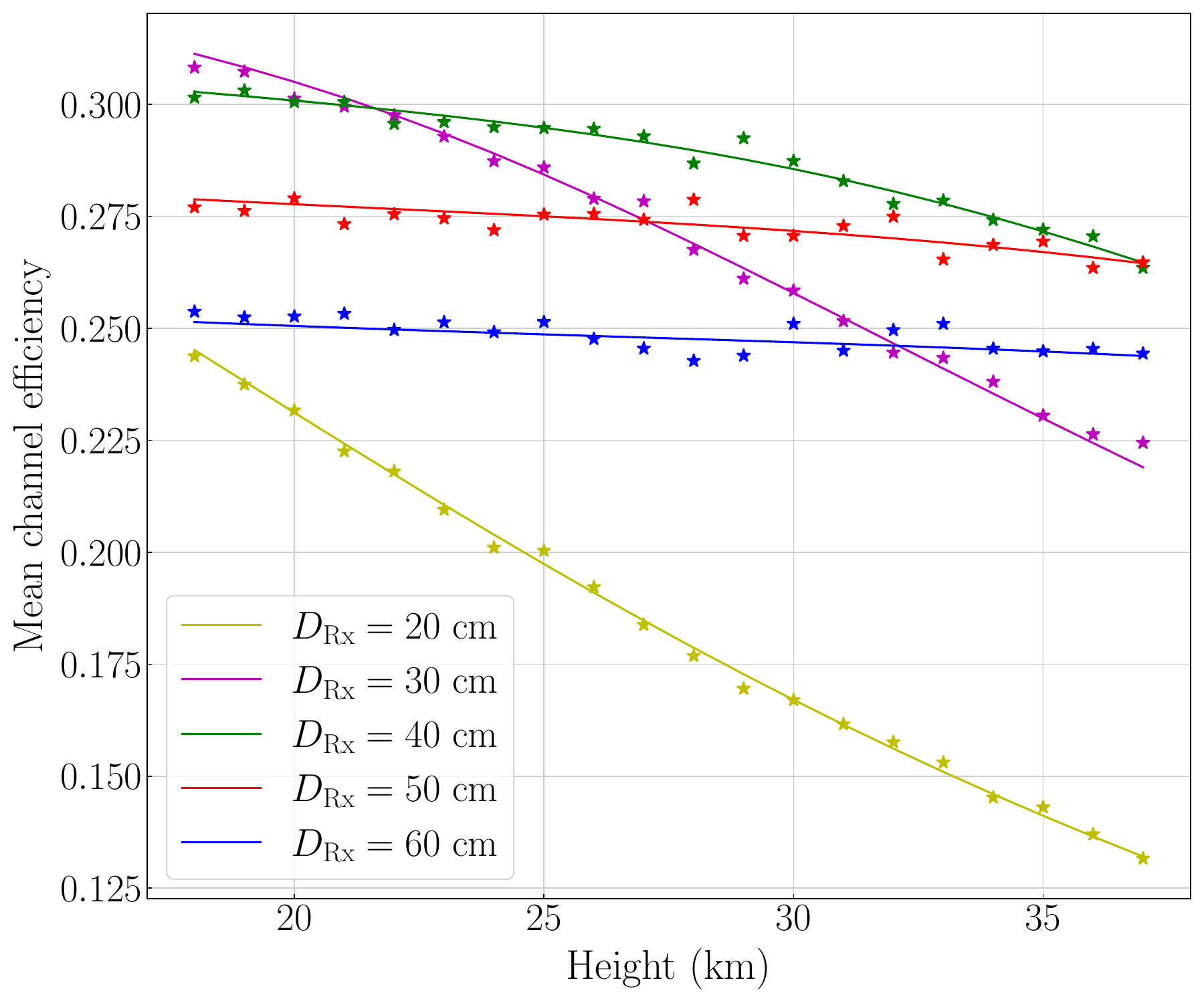}
    \caption{ }
    \label{fig:heightstudy}
 \end{subfigure}
\begin{subfigure}{.5\textwidth}
    \centering
    \includegraphics[width=\textwidth]{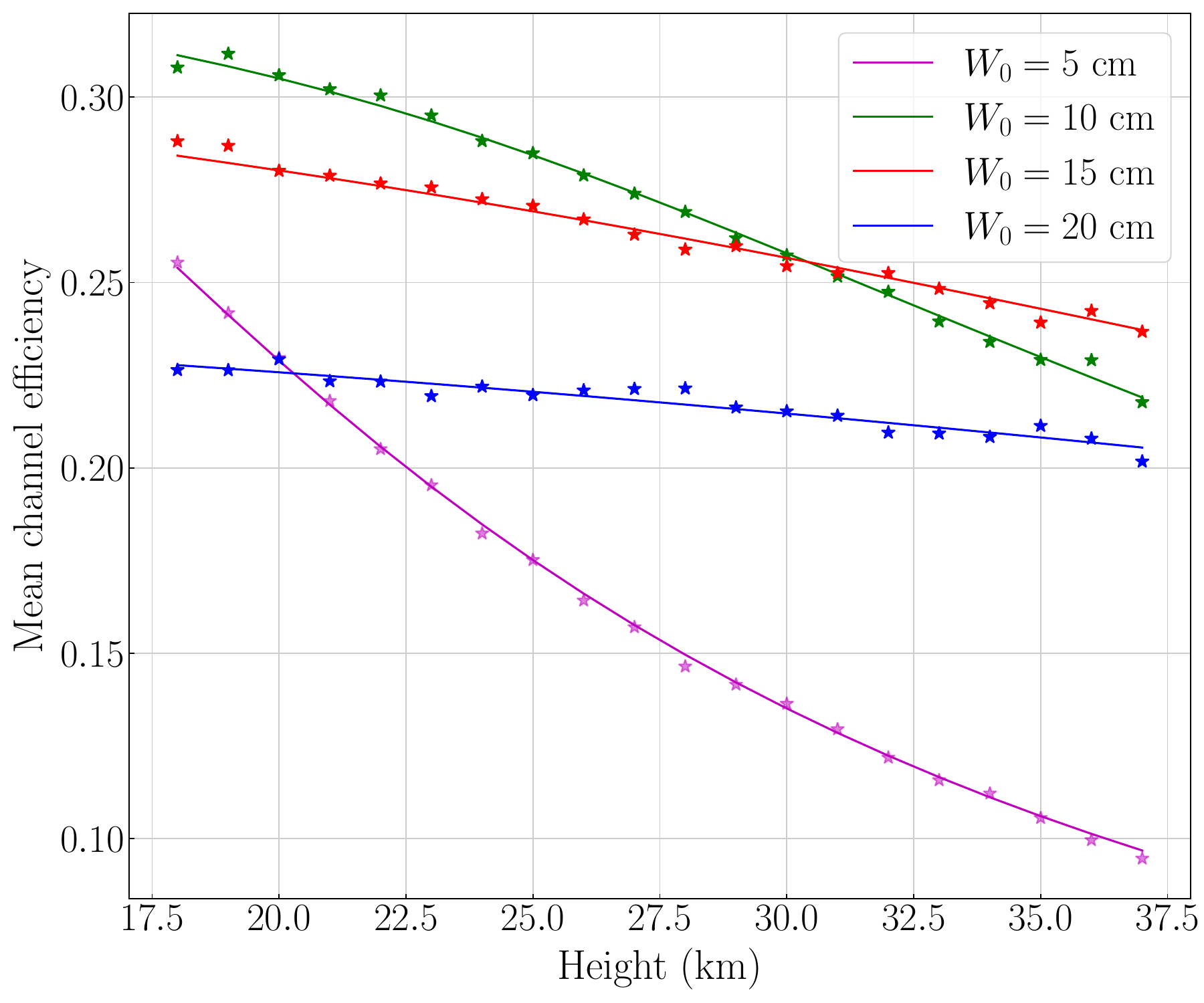}
    \caption{ }
    \label{fig:W0study}
\end{subfigure}
\caption{Theoretical (---) and simulated ($\star$) mean channel efficiency of the vertical downlink channel for different values of \textbf{(a)} the aperture of the receiving telescope $D_{\textrm{Rx}}$ with $W_0=10$ cm and \textbf{(b)} the initial beam waist radius $W_0$ with $D_{\textrm{Rx}}=40$ cm. The maximum radial index of correction by the AO system is $N_{\rm AO}=6$.}
\label{fig:downlink}
\end{figure}

In Fig.~\ref{fig:heightstudy}, we show the mean channel efficiency of the vertical channel as a function of the height for different values of the aperture of the receiving telescope $D_{\rm Rx}$, and in Fig.~\ref{fig:W0study} for different values of the initial beam waist radius $W_0$. The simulation was done by averaging over 45000 photons sent for each position of the balloon. The statistical error of the simulations is of the order of 0.01 and is not included in the plots to favor their readability.

With the plots in Fig.~\ref{fig:downlink}, we assess the correctness of the integration of the loss model in the Netsquid modules: simulated values are always within the statistical error range around the theoretical ones. A few interesting remarks can be mentioned about these first results. Namely, there are two types of trade-offs that influence the outcome, one between the mean collection efficiency and mean fiber-coupling efficiency with respect to the receiving aperture diameter (Fig.~\ref{fig:heightstudy}), and one regarding the dependence of both the beam size at the receiver and the wandering variance on the initial beam-waist radius $W_0$ (Fig.~\ref{fig:W0study}). 

These simulations suggest that a large receiving telescope aperture does not necessarily result in a higher channel efficiency. We indeed see in Fig.~\ref{fig:heightstudy} that a receiving telescope aperture diameter of 40 cm results in a higher channel efficiency than larger sizes. We also see that the channel efficiency is more affected by the length of the channel when $D_{\textrm{Rx}}$ is small. This is because the pointing error becomes more detrimental with distance. In Fig.~\ref{fig:W0study}, we see that transmitted beams with smaller initial waists are more impacted by the length of the channel, because their divergence is higher.

\begin{figure}[!ht]
    \centering
    \includegraphics[width=0.8\textwidth]{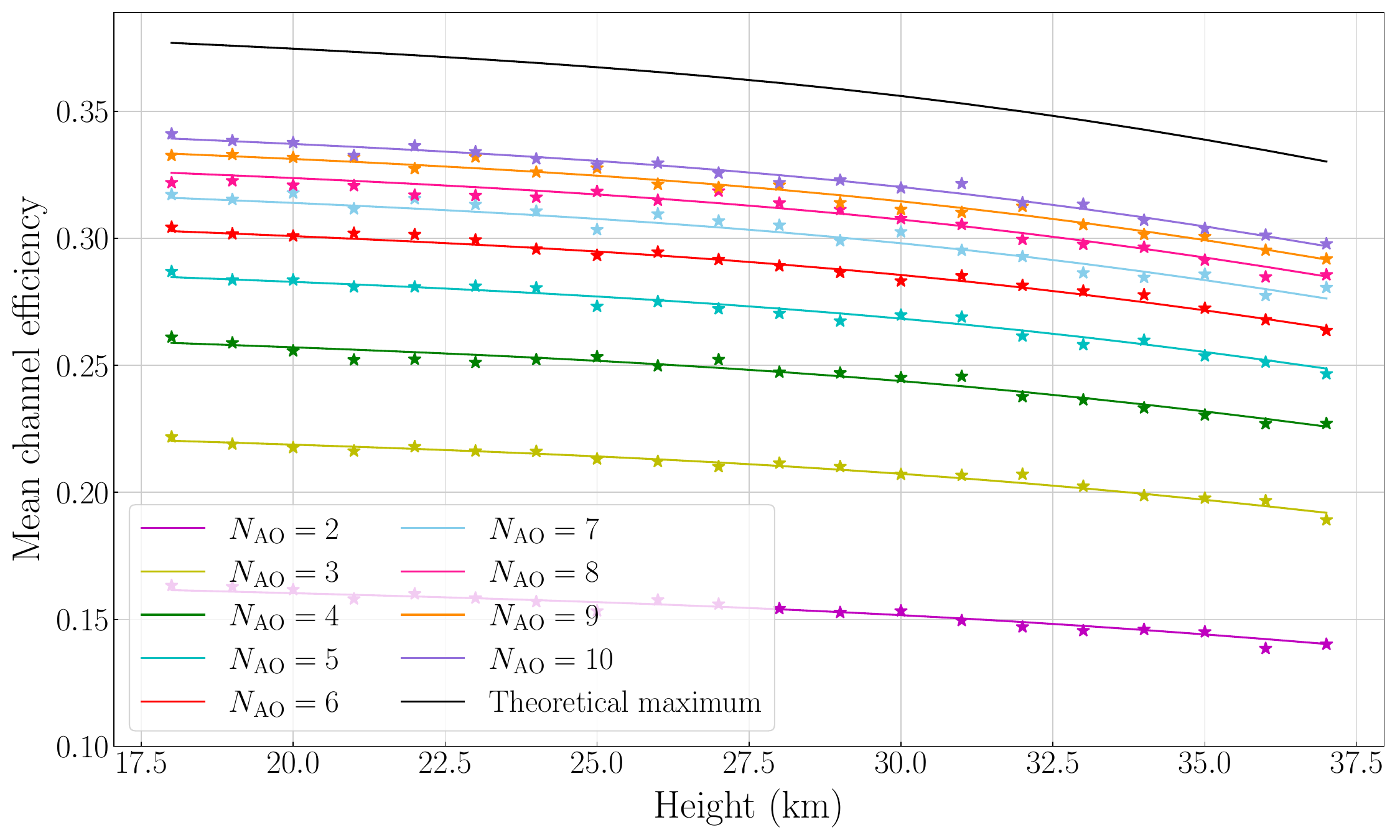}
    \caption{Theoretical (---) and simulated ($\star$) mean channel efficiency of the vertical downlink channel for different values of the order of correction of the adaptive optics system. The aperture of the receiving telescope is $D_{\textrm{Rx}}=40$ cm, and the initial beam waist is $W_0=10$ cm. The black curve represents the theoretical maximum with all the modes being perfectly corrected by the AO system.}
    \label{fig:AOstudy}
\end{figure}

The importance of adaptive optics and the influence of the order of correction can be seen in Fig.~\ref{fig:AOstudy}. As with satellite communication, adaptive optics greatly improve the channel efficiency, especially when correcting low-order aberrations, which are more detrimental to the quality of the received wavefront. 

\newpage

\begin{figure}[h!]
    \centering
    \includegraphics[width=0.8\textwidth]{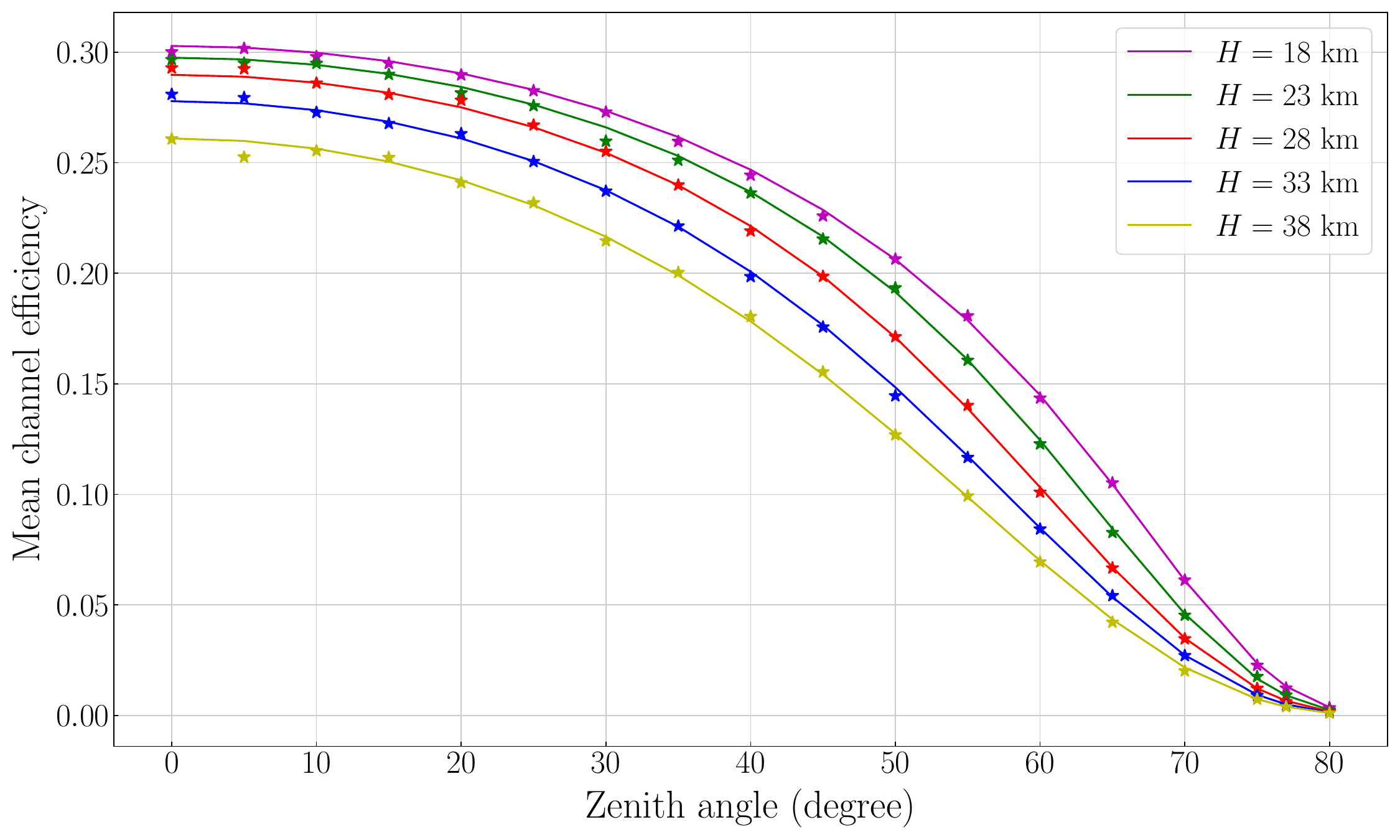}
    \caption{Theoretical (---) and simulated ($\star$) mean channel efficiency of the downlink channel as a function of the zenith angle for different heights of the balloon. The aperture of the receiving telescope is $D_{\textrm{Rx}}=40$ cm, the initial beam waist is $W_0=10$ cm and the maximum radial index of correction by the AO system is $N_{\rm AO}=6$. }
    \label{fig:zenithStudy}
\end{figure}

We then show in Fig.~\ref{fig:zenithStudy} the mean channel efficiency of the downlink channel for different values of the zenith angle and altitudes of the balloon. This gives an indication on how a system mounted on any aerial platform moving over a city during a period of time and continuously sending photons would perform. We also observe that, for small changes in the zenith angle ($<10$ degrees), the mean channel efficiency is mostly constant. This means that small deviations in the positions of an aerial platform above a city due, for example, to wind, would not significantly impact the channel efficiency as long as correct pointing is ensured. \\

\begin{figure}[!ht]
 \begin{subfigure}{.5\textwidth}
    \centering
    \includegraphics[width=\textwidth]{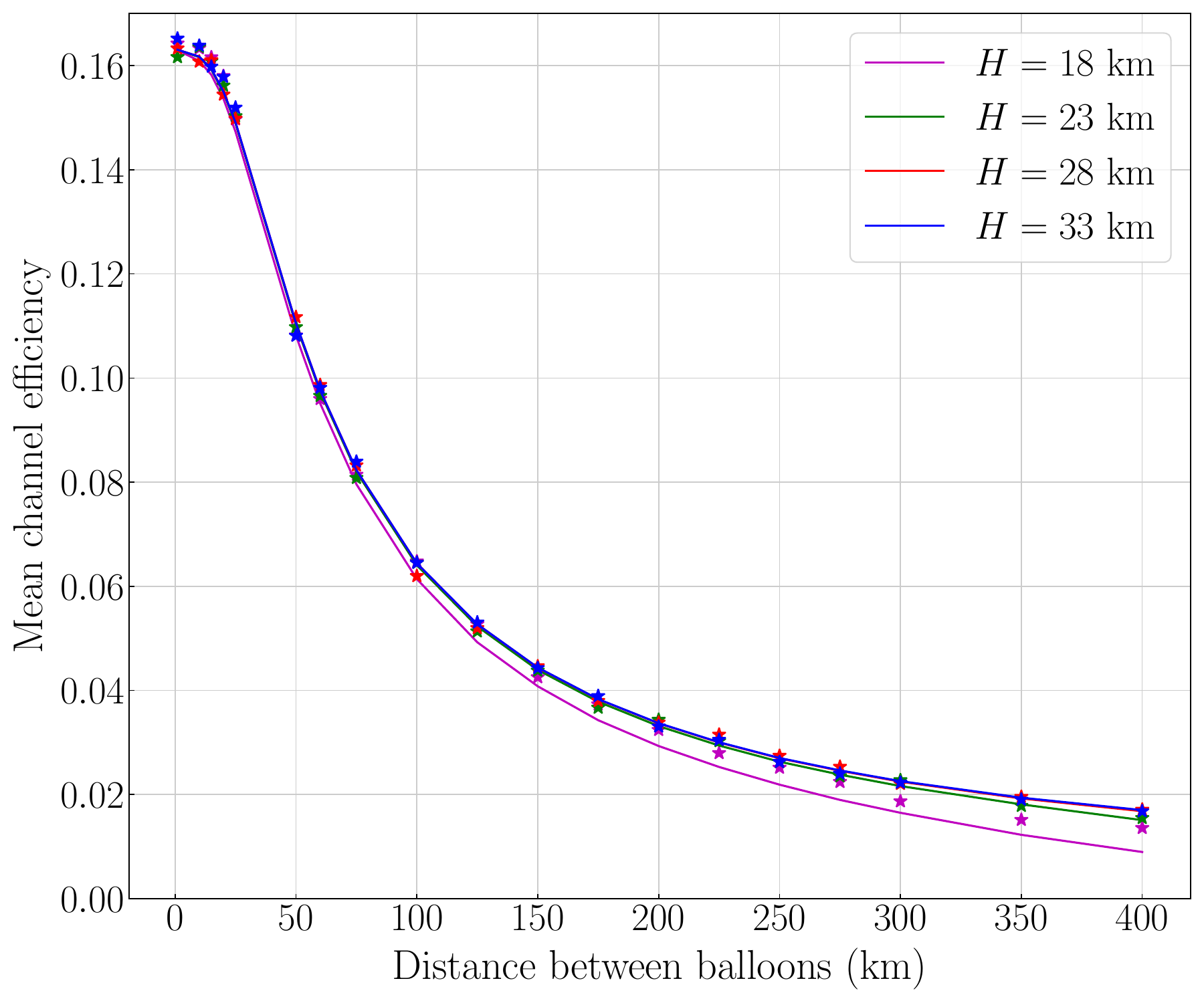}
    \caption{ }
    \label{fig:horizstudy}
 \end{subfigure}
\begin{subfigure}{.5\textwidth}
    \centering
    \includegraphics[width=\textwidth]{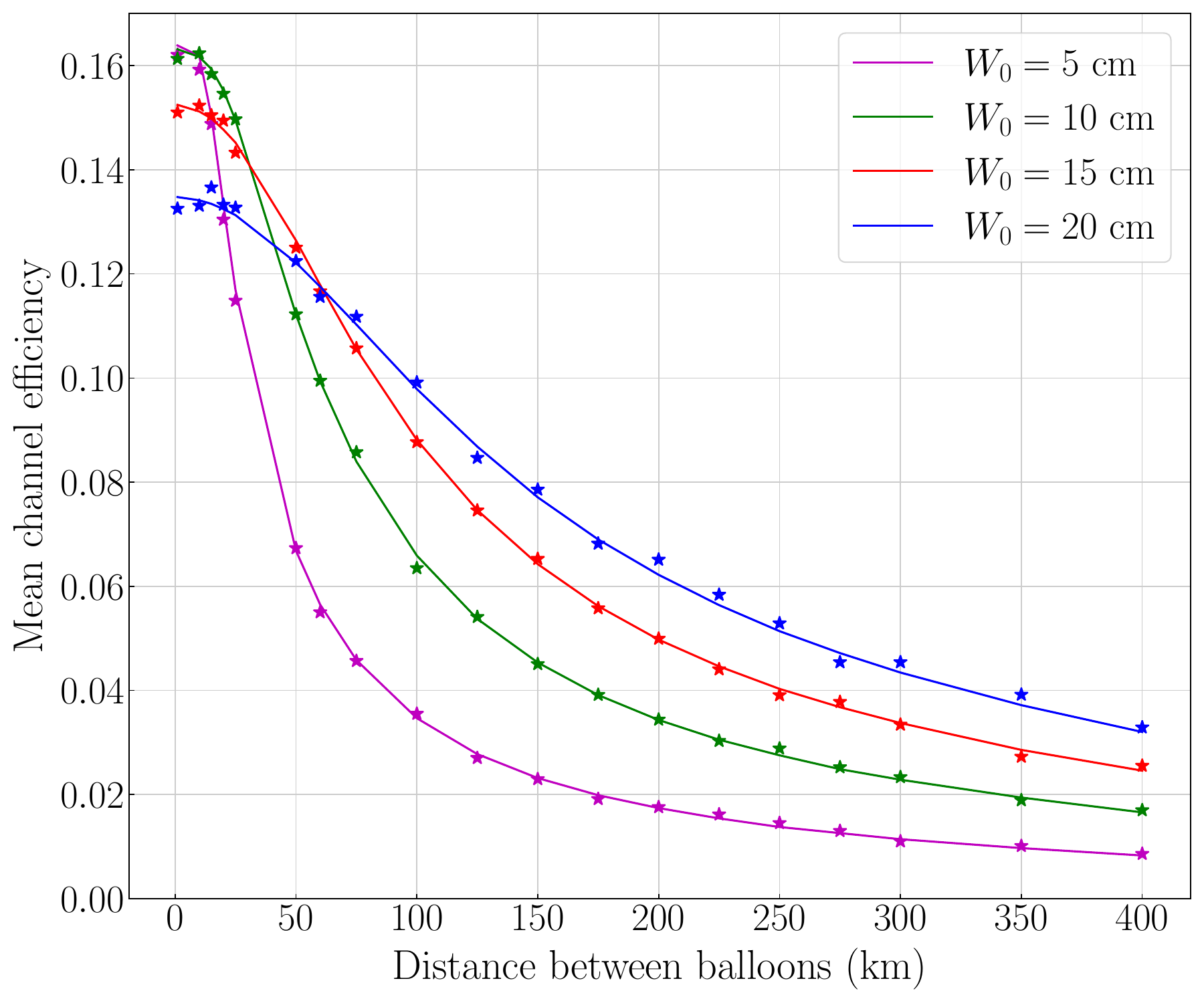}
    \caption{ }
    \label{fig:horizW0study}
\end{subfigure}
\caption{Theoretical (---) and simulated ($\star$) mean channel efficiency of the horizontal channel for different values of \textbf{(a)} the height of the balloons for $W_0=10$ cm and \textbf{(b)} the initial beam waist $W_0$ at a fixed height of 25 km. The aperture of the receiving telescope in the balloon is $D_{\textrm{Rx}}=30$ cm.}
\end{figure}

Regarding the horizontal channel, we show the mean channel efficiency between two balloons as a function of their distance, for different altitudes of the balloons in Fig.~\ref{fig:horizstudy}, and for different initial beam waists in Fig.~\ref{fig:horizW0study}. For the case of the horizontal link, in the calculation of the transmittance and the refractive index structure constant, we use the value of the minimal height of the path between the two balloons $h_{min}$ as it gives a lower bound on the total channel efficiency (see Fig.~\ref{fig:horiz}). We recall that, in the case where the receiver is in the balloon, there are no adaptive optics and the detectors in the balloon have a lower efficiency. In Fig.~\ref{fig:horizstudy}, we observe that the horizontal channel efficiency is  not significantly affected by the height of the balloon, as the atmospheric turbulence strength is mostly the same within the height ranges considered. In Fig.~\ref{fig:horizW0study} however, we observe a trade-off between the initial beam waist and the length of the channel. For horizontal channels over 75 km, it is better to use a larger initial beam waist. \\

\begin{figure}[!ht]
    \centering
    \includegraphics[width=0.8\textwidth]{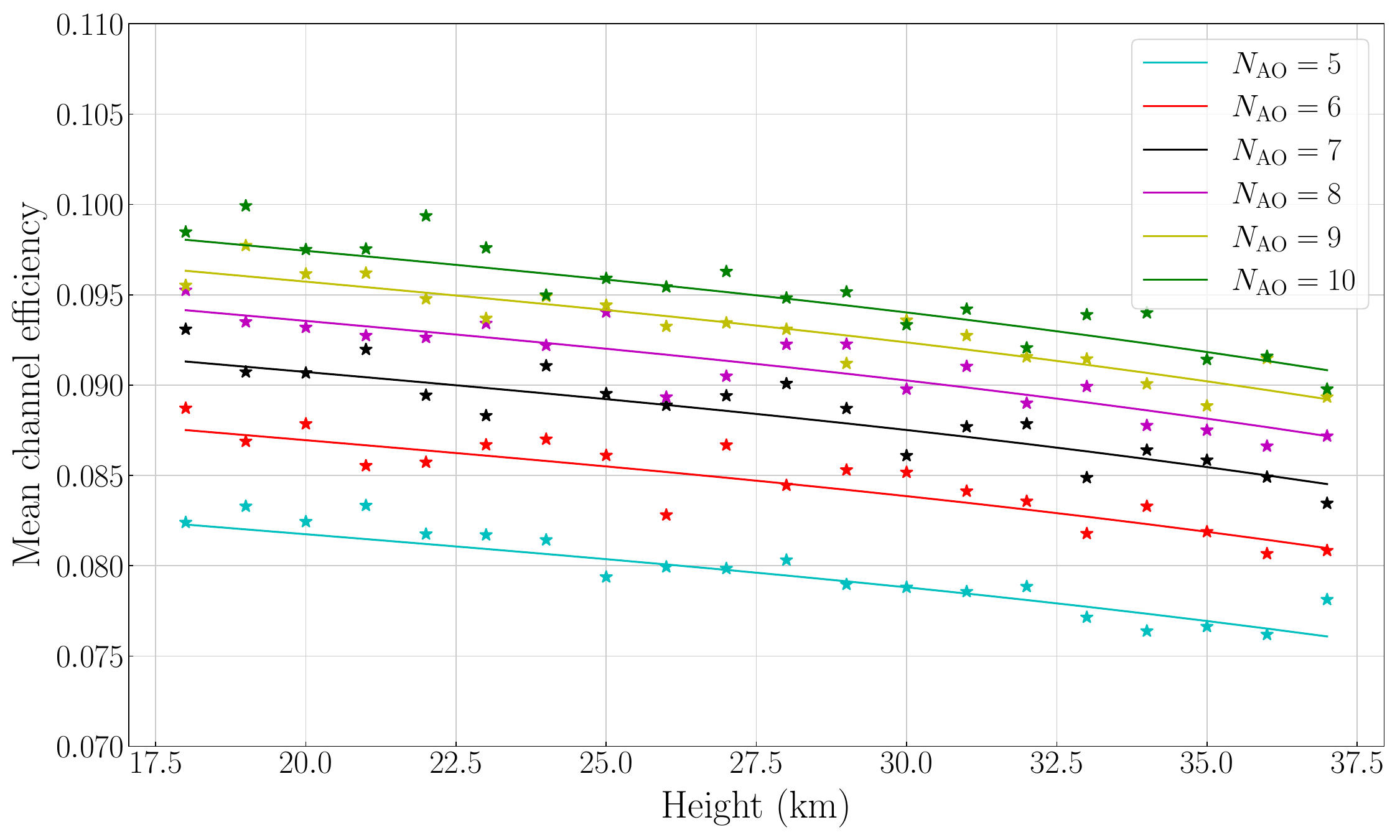}
    \caption{Theoretical (---) and simulated ($\star$) mean channel efficiency of the vertical uplink channel as a function of the height of the balloon for different values of the maximal level of correction of the AO system $N_{\rm AO}$. The aperture of the receiving telescope at the balloon is $D_{\textrm{Rx}}=30$ cm and the transmitted beam waist radius from the ground station is $W_0=20$ cm.}
    \label{fig:uplink}
\end{figure}

Finally, we take a look at the performance of the uplink channel to study the feasibility of ground-to-balloon quantum communication. In Fig.~\ref{fig:uplink}, we present the mean channel efficiency of the vertical uplink channel as a function of the height of the aerial platform. We consider a receiver aperture of $D_{\textrm{Rx}}=30$ cm, which is a realistic value based on the experimental realization of \cite{Wang2013}. Note that the fluctuations of the simulation results are more visible in this plot than in the previous ones because the range of values of the mean channel efficiency is small. Nonetheless, the simulation results are still within the same statistical error as before. 

We recall that our model for this channel is based on reciprocity with the downlink channel, as explained in Appendix~\ref{sec:uplink}. We can see that the mean channel efficiency is almost half compared to the vertical downlink channel. This is partly due to the less efficient detectors that are used in the aerial platform. We see however, that the performance of the uplink channel is positive. This means that we can, for example, devise a scheme with a balloon performing a bell-state measurement between two users, effectively creating a so-called quantum repeater between two nodes. We discuss this possibility in more detail in Sec.~\ref{sec:MDI}. \\

These simulations show promising results towards the feasibility of balloon-based quantum communication. Our simulation tool has also allowed us to study the interplay of different parameters of free-space quantum channels. We will now include balloons in a hybrid network architecture.

\subsection{QKD between users of a balloon-based quantum network}

In this section, we demonstrate how a balloon-based quantum network would perform in real-life scenarios. Following the method of~\cite{QCity,yehia2023connecting}, we compute the achievable secret key rate between two users of distant quantum cities. In this particular study, we choose to model a quantum network in Italy, with a first Qlient (Alice) based in Venezia (Venice), and another (Bob) based in Siena, wishing to generate a secret key. Via a fiber-based link, they connect to their local Qonnector, respectively based in Padova (Padua) and in Firenze (Florence). As explained in~\cite{QCity,yehia2023connecting}, connecting to a Qonnector is equivalent to connecting to the quantum network. We chose these particular cities for the purpose of providing a benchmark, but we emphasize that these are arbitrary choices and that our open-source library~\cite{Newgithub} can modularly test the efficiency of a QKD network between any cities.

\subsubsection{Parameters and figures of merit}

We compute the achievable secret key rate (SKR) between Alice and Bob with two different network architectures: one where each Qonnector has a balloon right above it, and one where a balloon is placed in the middle of the two cities. We also study two different QKD scenarios: a trusted node BB84-based~\cite{BB84} scenario and an untrusted node entanglement-based~\cite{Ekert,BBM92} scenario, showcasing the trade-off between rate and security level. To compute the SKR, we use the following formula~\cite{SecurityQKD, BB84proof,repeaterNV}
\begin{equation}
\label{eq:skr}
    \mathrm{SKR} = R\cdot[1 -  h(Q_x) -  h(Q_z)],
\end{equation}
where $R$ is the \textit{raw key rate} in bits per second, $Q_x$ and $Q_z$ are the \textit{qubit error rates} (QBER) in the two measurement bases and h is the binary entropy function. As a simplification, we will consider that the QBER is the same in both measurement bases, and set it at $Q_x=Q_z=4$\% \cite{picciariello2023intermodal}. The raw key rate is given by
\begin{equation}
\label{eq:rawkey}
    R= r_{\mathrm{source}} \cdot \mu \cdot \eta_{\mathrm{channel}}  ,  
\end{equation}
where $r_{\mathrm{source}}$ is the rate of the source, $\mu$ is the success probability of emitting a state (or the mean photon number per pulse in the case of weak coherent states), and $\eta_{\mathrm{channel}}$ is the mean channel efficiency. The parameter $\eta_{\mathrm{channel}}$, which includes the detection efficiency, is computed by our simulation module, as in the previous section, by looking at the ratio between the number of photons received over the number of photons sent over a given channel. \\

In all that follows, we will consider that the parameters of the free-space channel are fixed and correspond to the values given in Table~\ref{tab:QKDparameters}. The other parameters are listed in Table~\ref{tab:baselineparameters}.  
\begin{table}[!ht]
    \centering
\begin{tabular}{l|c|l}
    Symbol & Value & Description  \\ \hline
    $H$ & $35$ km & Height of the balloons\\
    $W_{0,\textrm{ ground}}$ & $20$ cm & Initial beam waist radius on the ground \\ 
    $W_{0, \textrm{ balloon}}$ & $10$ cm & Initial beam waist radius on the balloon \\ 
    $D_{\textrm{Rx, ground}}$ & $40$ cm  & Diameter of the receiving telescope on the ground  \\
    $D_{\textrm{Rx, balloon}}$ & $30$ cm  &Diameter of the receiving telescope in the balloon  \\
    $N_{\rm AO}$ & $6$ & Maximum radial order corrected by AO on the ground \\
    $r_{\mathrm{source}}$ & $80$ MHz & Rate of the source \\
    $\mu$ & $0.01$ & Mean photon number per pulse \\
    $Q_x,Q_z$ & $4\%$ & QBER in the two measurement basis \\
\end{tabular}
    \caption{Parameters for the QKD simulations.}
    \label{tab:QKDparameters}
\end{table}

As explained in Sec.~\ref{sec:setup}, there are a few conditions imposed by our model to ensure the accuracy of the simulation results. These conditions are applicable in the case of the downlink and uplink channels, and become more stringent for slanted propagation. We show in Appendix~\ref{sec:verifassumption} that these conditions are met in the context of the Italian quantum network that we present in the following sections. On top of the two QKD scenarios, we also study the possibility of using a balloon as a middle node to perform measurement-device independent QKD (MDI-QKD)~\cite{MDIQKD} between two nodes, which also gives some perspective on using balloons as repeater nodes.

\subsubsection{Trusted node scenario}
\label{sec:trustednodescenario}
\begin{figure}[!ht]
 \begin{subfigure}{\textwidth}
    \centering
    \includegraphics[width=12cm]{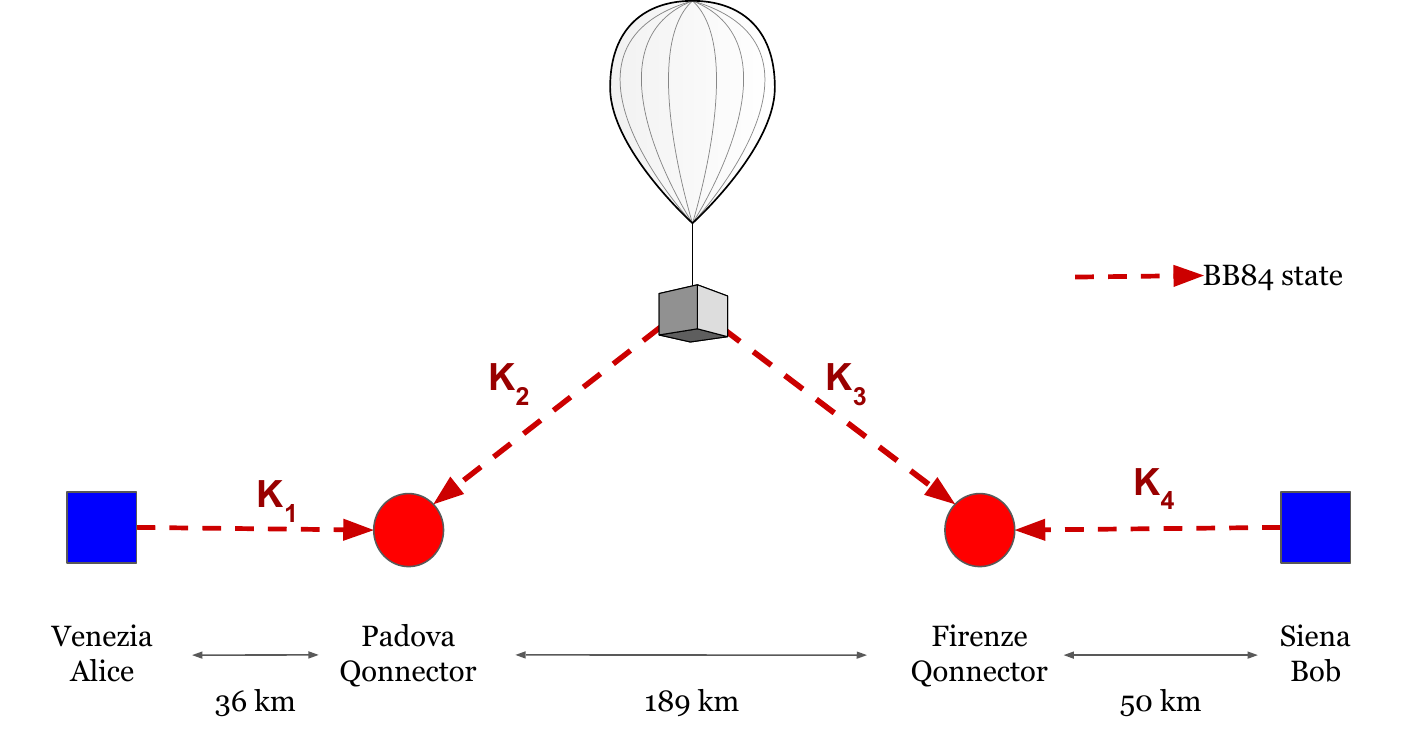}
    \caption{}
    \label{fig:trustedscenar1}
 \end{subfigure}
\begin{subfigure}{\textwidth}
    \centering
    \includegraphics[width=12cm]{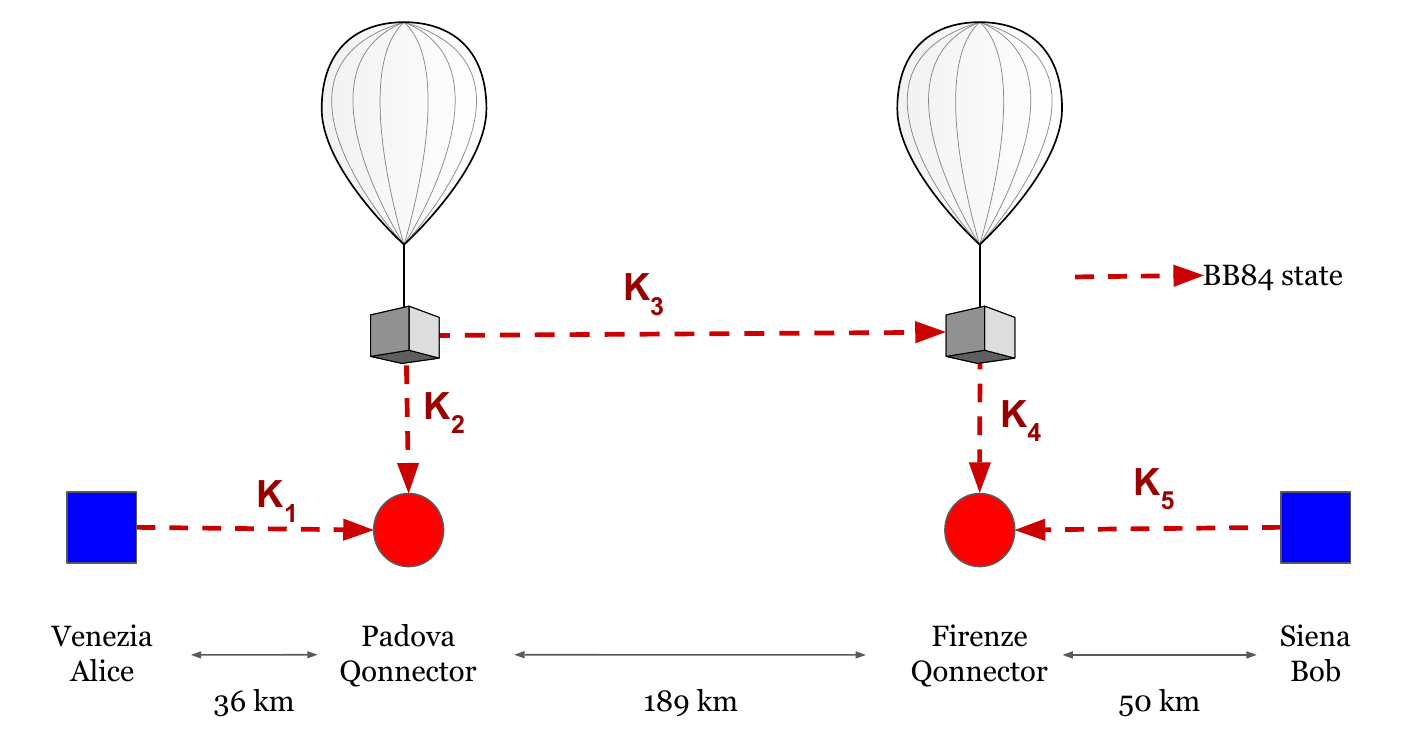}
    \caption{}
    \label{fig:trustedscenar2}
\end{subfigure}
\caption{Trusted node scenarios to generate a key between Alice and Bob: \textbf{(a)} a balloon is placed in the middle between the two Quantum cities composed of a Qonnector and a Qlient and \textbf{(b)} a balloon is placed above each Qonnector. In both cases, a key is generated in parallel on all the sublinks between Alice and Bob using the BB84 protocol, and then $K_1$ is transmitted classically from Alice to Bob through each sublink using the other keys.}
\label{fig:trustedscenar}
\end{figure}

In the first scenario, the two users wishing to share a key, Alice and Bob, trust the nodes of the network between them. The resulting architectures are showed in Fig.~\ref{fig:trustedscenar}, where we also show the bird flight distance between the cities considered. Each node performs the BB84 protocol with its neighbour, generating secret keys $\{K_i\}_{i=1}^{5}$ for each sub-link along the path between Alice and Bob. The key $K_1$ is then transferred classically from Alice to Bob. This can be done securely by using, for example, one-time pad protocols. More explicitly, in the example architecture of Fig~\ref{fig:trustedscenar1}, $K_1$ is encrypted using $K_2$ and sent securely by the Padova Qonnector to the balloon, which encrypts it using $K_3$ and sends it to the Florence Qonnector, which finally encrypts it using $K_4$ and sends it to Bob. 

In Table~\ref{tab:trusted}, we show the secret key rate for each of the sublinks between the two Qlients, computed by combining equations~\eqref{eq:skr} and \eqref{eq:rawkey}, with $\eta_{\mathrm{channel}}$ given by our simulation tool. More precisely, we averaged over 20 repetitions of sending over 150000 photons on each channel and counting the number of photons received. Assuming that each link is performing the BB84 protocol in parallel, the secret key rate achievable between Alice and Bob is given by the minimum rate achievable in the sub-links along the path between them. For the specific network considered, the most optimal architecture is the one from Fig~\ref{fig:trustedscenar2}, with a balloon above each city. Indeed, a balloon placed between these two cities at a 35 km height has a zenith angle of almost 70 degrees with respect to the Qonnectors which, as we can see from the simulation in Fig.~\ref{fig:zenithStudy}, causes a low channel efficiency. Having a horizontal balloon-to-balloon free-space link minimizes the propagation into the densest part of the atmosphere and allows the two Qlients to achieve a larger secret key rate than using slant links.

\begin{table}[!ht]
    \centering
    \begin{tabular}{|c|c|}
    \hline
         Link  & Secret key rate (kbit per second)\\ \hline
         Slant link Balloon $->$ Qonnector & $2.21 \pm 0.144$\\
         Vertical downlink Balloon  $->$ Qonnector & $112.01 \pm 0.735$ \\
         Horizontal link Balloon $->$ Balloon & $24.65 \pm 0.451$\\ \hline
         Alice $->$ Qonnector & $70.71 \pm 0.685$  \\
         Bob $->$ Qonnector & $39.85 \pm 0.371$   \\ \hline
 
    \end{tabular}
    \caption{Performance of the BB84 protocol between all nodes. The first three rows correspond to the different free-space channels considered, and the other rows correspond to the fiber links between each Qlient and their Qonnectors. }
    \label{tab:trusted}
\end{table}

\subsubsection{Untrusted node scenario}
\begin{figure}[!ht]
    \centering
    \includegraphics[width=12cm]{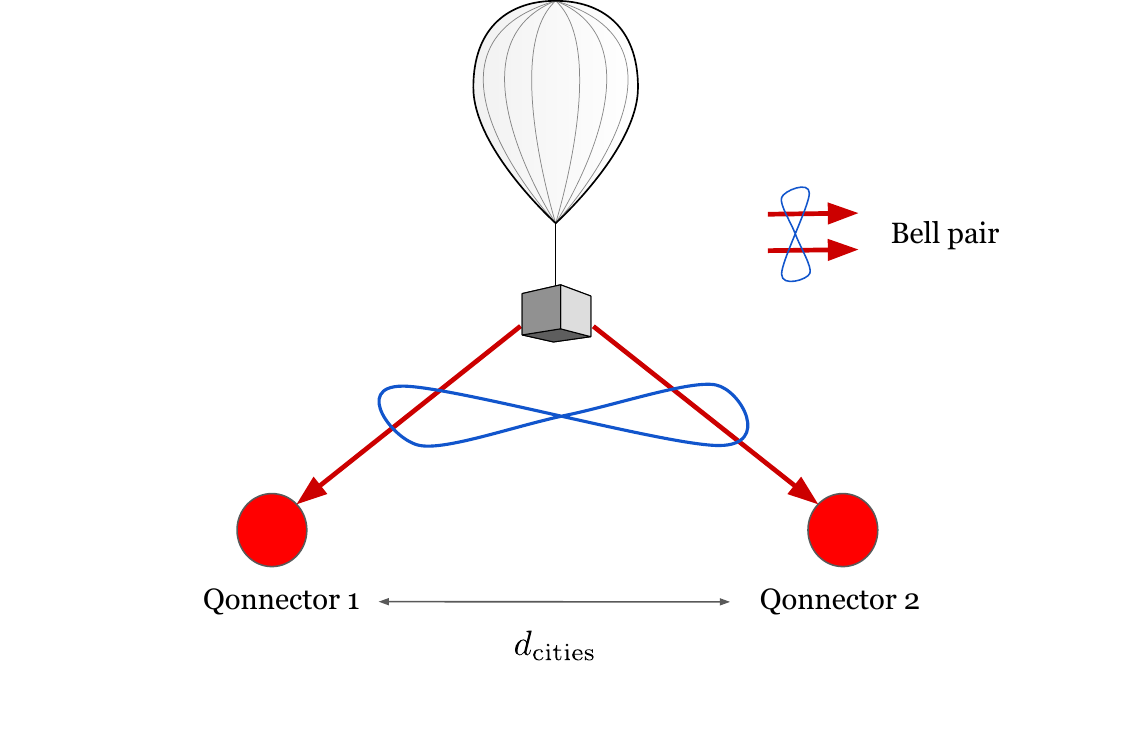}
    \caption{Entanglement-based QKD between two ground stations separated by a distance $d_{\rm cities}$. The balloon, placed above the middle point between the two Qonnectors, shares photonic Bell pairs through free-space channels.}
    \label{fig:entgeneral}
\end{figure}

In the second scenario, the two users do not make any assumption on the nodes between them and only trust their measurement devices. This is possible by using entanglement-based QKD, in which Bell pairs are sent from a middle node to Alice and Bob. These Bell pairs can be created using, for example, spontaneous parametric down conversion (SPDC) in a non-linear crystal. By sharing some outputs of their measurements, Alice and Bob are able to certify that they indeed received Bell pairs, and can exploit the rest of their outputs to generate a secret key. We first study the possibility of using the balloon as a middle node sending Bell pairs to two ground stations, as shown in Fig.~\ref{fig:entgeneral}, and then apply it in the Italian network architecture presented above.

\begin{figure}[!ht]
    \centering
    \includegraphics[width=0.8\linewidth]{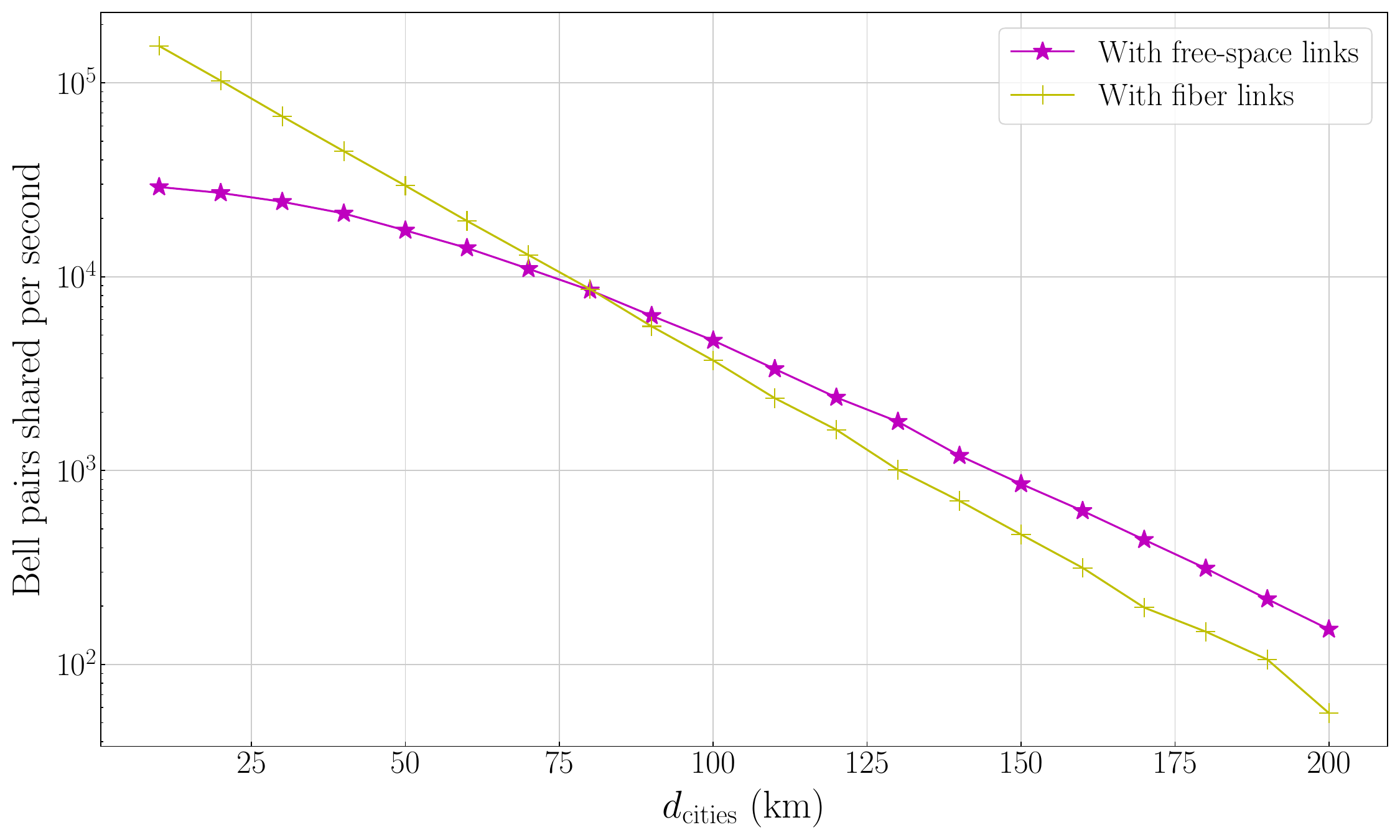}
    \caption{Comparison of the number of Bell pairs shared per second between the two Qonnectors (in log scale) as a function of the distance between the cities, using free-space links and a balloon as a middle node ($\star$), and using fiber links and a ground-based station as a middle node (+).}
    \label{fig:EPRfibervsfree}
\end{figure}

In Fig.~\ref{fig:EPRfibervsfree}, we compare the number of Bell pairs shared between the two cities using either a balloon as a middle node, or a ground-based middle node. In the first case, the two photons from the Bell pair are sent through free-space channels as modeled in this work, while in the latter case they are sent through fiber channels. Each point is computed by using Eq.~\eqref{eq:rawkey}, with $\eta_{\mathrm{channel}}$ obtained by counting the number of times where both photons of a Bell pair sent from the middle node arrive at the two Qonnectors. Since a successful Bell pair transmission happens with a low probability, the sending of around 760000 Bell pairs was simulated to get relevant statistics, which required a longer simulation time. We identify a clear crossing of the curves around 80 km, meaning that it is more efficient to use a balloon node and free-space links for sharing Bell pairs between two nodes separated by distances over 80 km. \\

\begin{figure}[!ht]
 \begin{subfigure}{\textwidth}
    \centering
    \includegraphics[width=12cm]{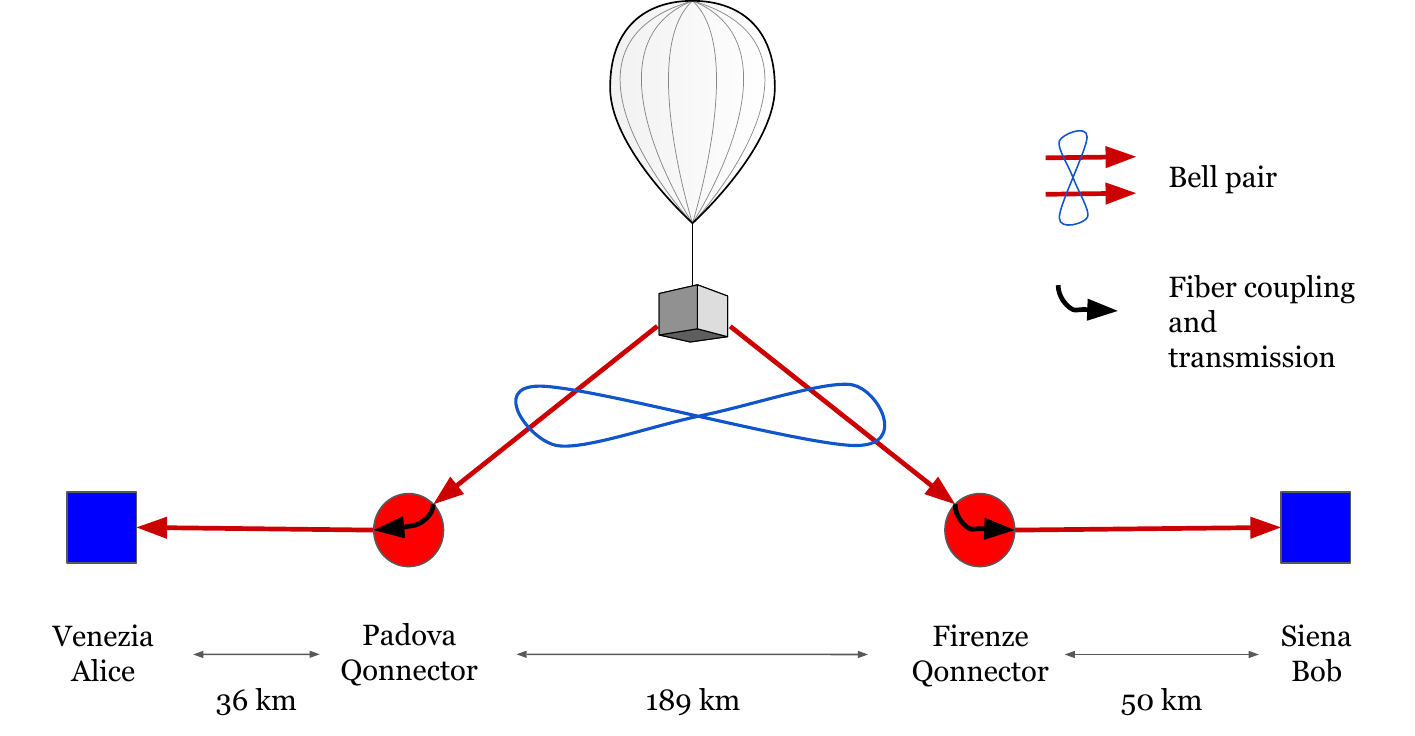}
    \caption{}
    \label{fig:untrustedscenar1}
 \end{subfigure}
\begin{subfigure}{\textwidth}
    \centering
    \includegraphics[width=12cm]{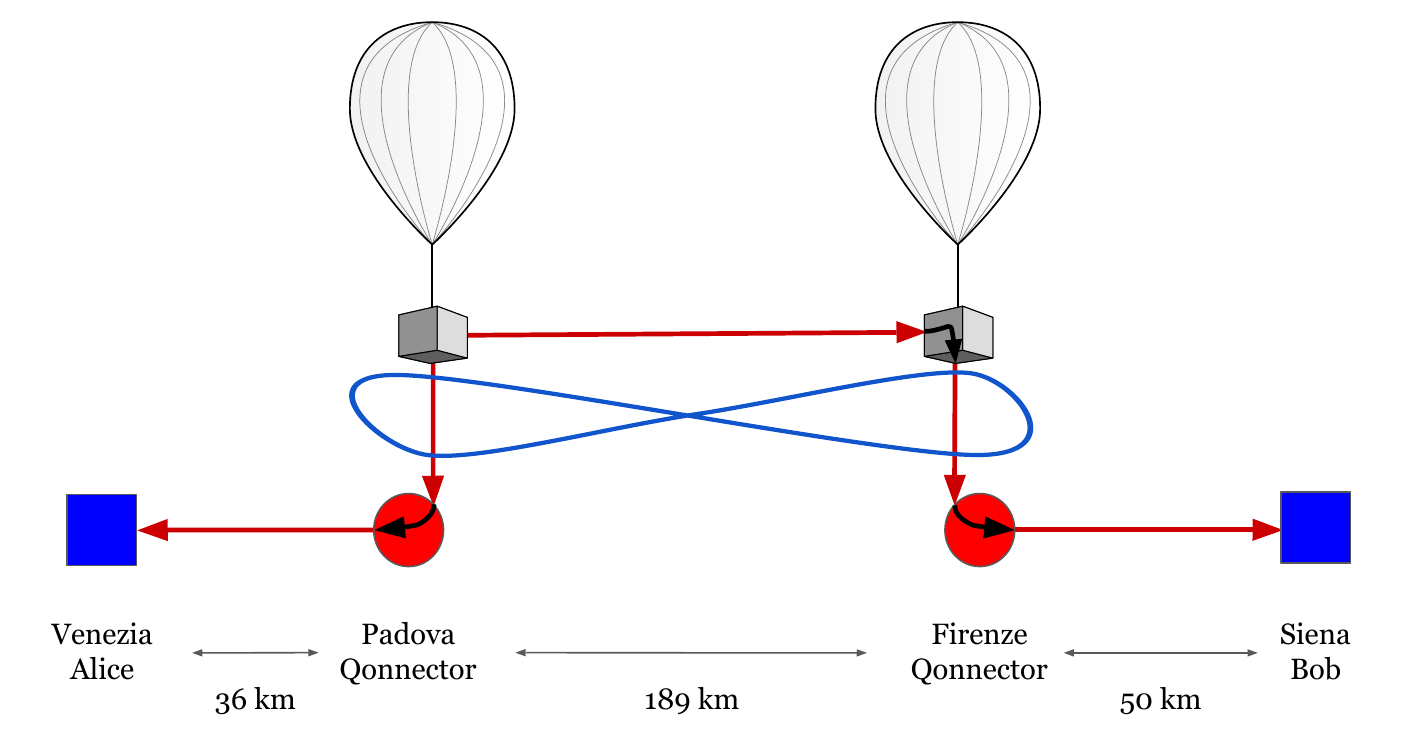}
    \caption{}
    \label{fig:untrustedscenar2}
\end{subfigure}
\caption{Untrusted node scenarios to generate a key directly between Alice and Bob, using the entanglement-based QKD scenario: \textbf{(a)} a balloon, between the cities, creates and sends photonic Bell pairs to the two Qonnectors who transmit each photon to each Qlient, and \textbf{(b)} a balloon is placed above each Qonnector. One of them creates and sends Bell pairs that are then routed to each Qlient.}
\label{fig:untrustedscenar}
\end{figure}

In the setting of the Italian network, we now use a balloon, placed between the two cities or above one of them, to create and send Bell pairs. The two entangled photons are then routed to the Qlients through each node along the path. We show the two resulting architectures in Fig.~\ref{fig:untrustedscenar}. Note that we do not model how the routing is performed in the nodes between Bell pair generation and measurement. In the ground station nodes, this can be done by coupling the photon arriving from each of the balloons in a fiber going directly to Alice or Bob. In the balloon node, as in Fig.~\ref{fig:untrustedscenar2}, we assume that one balloon is able to couple a photon arriving from the other one into a fiber and transmit it via a telescope to the ground station below, without measuring it. While this is out of the scope of this work, we acknowledge that this might prove experimentally challenging.

In Table~\ref{tab:untrusted}, we show the number of Bell pairs received by Alice and Bob, as well as the secret key rate for the two scenarios considered. Since they are quite far from the ground stations, the probability that both photons from the Bell pairs successfully arrive from the balloon to the two Qlients, through the free-space and fiber channels, is small. This explains the relatively low number of Bell pairs received by Alice and Bob. We note, however, that the scenario using a horizontal free-space link (Fig.~\ref{fig:untrustedscenar2}), is again the most efficient one, for the same reason that we minimize the quantum communication over a dense part of the atmosphere. While the secret key rate is significantly lower than in the trusted node scenario of Sec.~\ref{sec:trustednodescenario}, we emphasize that we do not assume any trust in the nodes between the end users in this case. We recover a well-known trade-off between security assumptions and efficiency in communication protocols.

\begin{table}[!ht]
    \centering
    \begin{tabular}{|c|c|c|}
    \hline
         Scenario  & Bell pairs received by the two Qlients per second & Secret key rate (bit/s)\\ \hline
         Scenario 1 (Fig~\ref{fig:untrustedscenar1}) & $4 \pm 2$ & $2 \pm 1$\\
         Scenario 2 (Fig~\ref{fig:untrustedscenar2}) & $ 143 \pm 24 $ & $75 \pm 12$ \\ \hline
 
    \end{tabular}
    \caption{Performance of the entanglement-based protocol between Alice and Bob for the two scenarios considered.}
    \label{tab:untrusted}
\end{table}

\subsubsection{Measurement Device-Independent QKD}
\label{sec:MDI}
MDI-QKD~\cite{MDIQKD} is a scheme in which the two parties, Alice and Bob, both produce a single photon or a weak coherent state and send it to a third party, placed in the middle, where a Bell-state measurement (BSM) is performed. The protocol is secured against eavesdroppers or malicious middle nodes, since correlations are measured instead of actual bit values. MDI-QKD has gained a lot of popularity in recent years due to experimental realizations at large distances~\cite{MDIChina,Cao_2020}.\\

\begin{figure}[!ht]
    \centering
    \includegraphics[width=11.5cm]{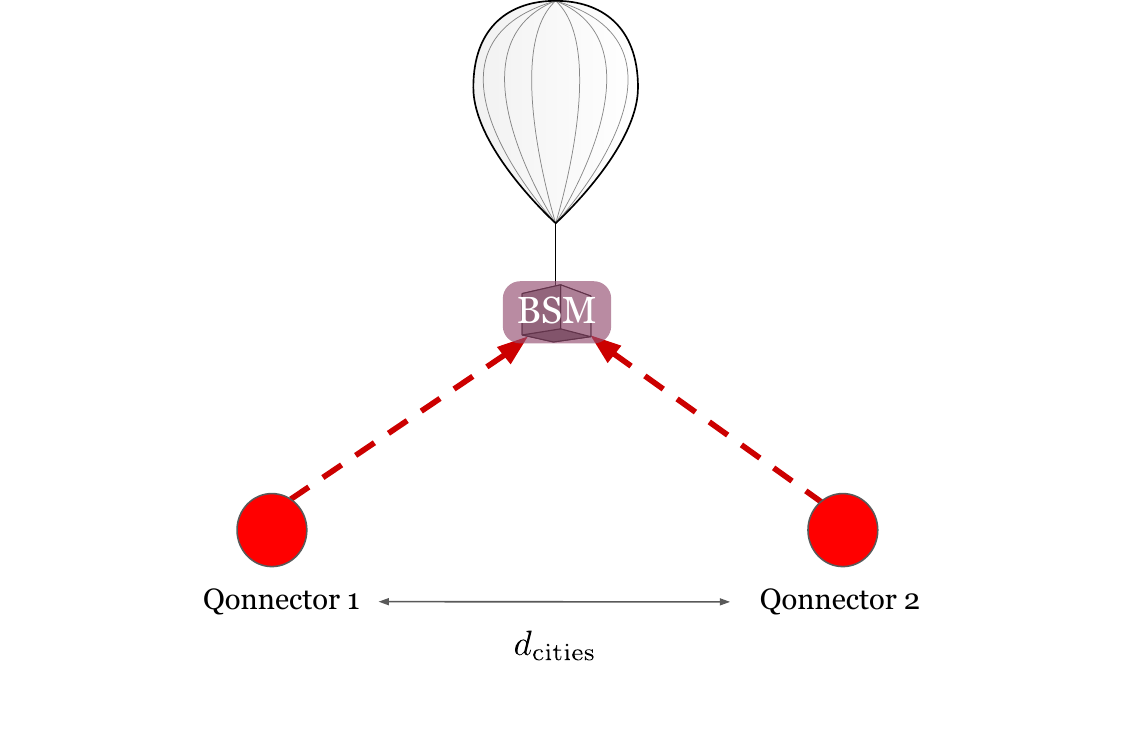}
    \caption{MDI-QKD between two ground stations separated by a distance $d_{\rm cities}$. The balloon, placed above the middle point between the two Qonnectors, receives single photons from the ground stations, performs a Bell-state measurement (BSM) and sends the outcome to the Qonnectors.}
    \label{fig:MDI}
\end{figure}

Here, we explore the possibility of using a balloon as a middle node in an MDI-QKD experiment. In this setting, the two ground stations send single photons through an uplink free-space channel towards a balloon placed above the middle point between them (see Fig.~\ref{fig:MDI}). The balloon performs a BSM and sends the measurement outcomes to both Alice and Bob. The success probability of this BSM, $p_{\rm BSM}$, is the main bottleneck of this protocol. Photonic BSM, without additional ancilla qubits, can only distinguish two of the four Bell-States~\cite{BSMwancilla1,BSMwancilla2}, which upper bounds $p_{\rm BSM}$ to 0.5. On top of that, two detectors in the balloon need to click. Hence, we set $p_{\rm BSM}$ to 
\begin{equation}
p_{\rm BSM} = 0.5 \cdot  p_{\rm det}^{\rm SPAD} \cdot  p_{\rm det}^{\rm SPAD} = 0.03125  
\end{equation}

\begin{figure}[!ht]
    \centering
    \includegraphics[width=0.8\linewidth]{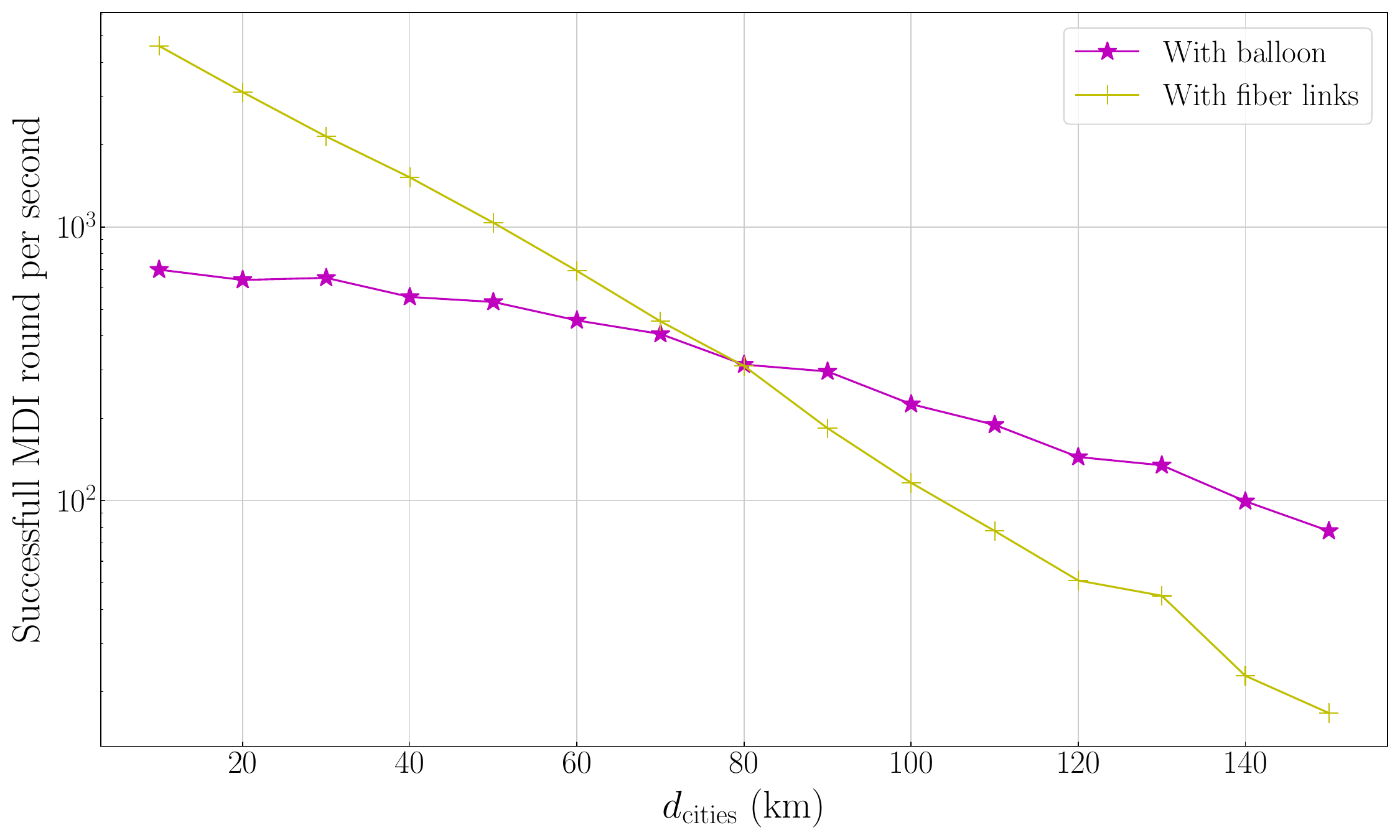}
    \caption{Comparison of the number of successful MDI-QKD rounds per second (in log scale) as a function of the distance between the cities, using free-space links and a balloon as a middle node ($\star$), or using fiber links and a ground-based station as a middle node (+).}
    \label{fig:MDIfiberVsfree}
\end{figure}

In Fig.~\ref{fig:MDIfiberVsfree}, we show the number of successful MDI-QKD rounds per second as a function of the distance between the cities, using either a balloon, as in Fig.~\ref{fig:MDI}, or fiber links and a third party placed in the middle between the two parties. For this simulation we set $N_{\rm AO} = 10$ and we simulate more than 900000 rounds in order to get relevant statistics. The number of successful MDI-QKD rounds corresponds to the raw key rate of the protocol. We see that, as with entanglement-based QKD, there is a clear cross of the curves around $d_{\rm cities} = 80$ km, meaning that it is more advantageous to use a balloon over this distance. We note that implementations of MDI-QKD involve additional experimental challenges, such as phase stabilization and numerical methods to compute the secret key rate, that we do not include in this study.

Due to its low success rate, the simulation of MDI-QKD between Qlients of the Italian network presented in this section yields a very small rate. Indeed, adding fiber losses on top of the uplink free-space losses, combined with the low success probability of the BSM, results in a very low channel efficiency. The simulation time to extract a rate with satisfactory accuracy increases the computational complexity. Thus, we leave a more complete study of balloon-based MDI-QKD for future work. We emphasize here on the proof-of-concept nature of the results from Fig.~\ref{fig:MDIfiberVsfree}, showing that balloons could, in principle, be used for MDI-QKD between two cities separated by a few hundred kilometers.\\

This simulation also showcases the possibility to use balloons as quantum repeater nodes ~\cite{Repeater1,Repeater2,Repeater3}. Indeed, if the Qonnector nodes are equipped with quantum memories that are entangled with the photon sent to the middle node, the BSM acts as a swapping operation that results in entanglement generation between the two Qonnectors, as done for example in~\cite{repeaterNV,coopmans2021netsquid}. Modelling such quantum memories is however a challenging task that is out of the scope of this paper.

\section{Conclusion}
Long-distance communication is the main bottleneck for the establishment of a global quantum network, as fiber communication is efficient up to a few tens of kilometers. Current visions involve the use of free-space channels to overcome this issue, in particular using satellites to connect different cities. In this work, we have studied the feasibility of a balloon-based quantum network through simulation of realistic architectures including both free-space and fiber links, showing that balloons, or other aerial platforms, are a valid alternative to satellites for long-distance communication. Although they require more maintenance than satellites, balloons are easier and a lot cheaper to launch. They are also more flexible and can be visible from a ground station for a longer period of time compared to a satellite, whose usable window is on the order of a few minutes per day for LEO satellites.

We have modeled horizontal, downlink and uplink free-space channels for aerial platforms placed at an altitude of 18 to 38 km, taking into account the statistical nature of the receiver collection efficiency and single-mode fiber coupling efficiency in the presence of adaptive optics. The latter is particularly useful for systems operating under daylight conditions at 1550 nm. We generalized the results of previous works and made the models applicable to a variety of atmospheric conditions ranging from weak to strong, depending on the distance and zenith angle in the case of slant propagation. The developed software is open-source and can be used both to estimate average free-space channel efficiencies or sample from the statistical distributions for integration in the Netsquid network simulator.

Using these models, we have shown the feasibility of balloon-based quantum communication. We have analyzed the trade-off between the receiver aperture and the initial beam waist and asserted the importance of adaptive optics in the ground stations. We have also shown that small deviations in the position of the balloons do not significantly affect the channel efficiency. We have benchmarked this configuration by estimating QKD secret key rates for different architectures and security assumptions in a specific instance of an Italian quantum network. These key rates are comparable to or better than the reported key rates for satellite communication. Following the approach started in \cite{QCity,yehia2023connecting}, these results provide a fundamental tool for experimentalists and quantum network engineers to study the best configuration for the future quantum internet. \\

Future works will include quantum memories into our model, allowing for the simulation of more complex network protocols, as well as the incorporation of a noise model in the free-space channels. A more comprehensive network architecture integrating various types of nodes such as drones, satellites, balloons, and ground stations, will be developed to simulate a complete quantum internet architecture. Another interesting outlook would be the study of delegated quantum computation \cite{BroadbentDelegated} by including parties equipped with quantum computers in our simulator.

Additionally, improvements could be made to address some of the limitations of the free-space model. For instance, applying a more generalized collection efficiency PDF beyond the weak beam wandering assumption, such as the one in \cite{vasylyev2018theory}, would broaden its applicability. In the case of the fiber-coupling efficiency PDF, it would be beneficial to create an alternative model that does not rely on having a very small residual phase variance after adaptive optics correction, as seen in efforts such as \cite{lognoneOptimizationHighData2023}. In that way, it would be possible to simulate a larger variety of adaptive optics systems, including the ones that are less efficient. Lastly, deriving the appropriate conversion matrix to transform the annular Zernike basis coefficients into an orthonormal set for the weighted pupil would be a step in making the model even more accurate.

\section*{Acknowledgements}

The authors thank Francesco Picciariello for allowing us to use part of his code in ours, Jean-Marc Conan for fruitful discussions, as well as Antonio Ac\'in, Eleni Diamanti, Giuseppe Vallone and Paolo Villoresi for their feedback and constructive insights on the manuscript.

I.~Karakosta-Amarantidou acknowledges funding from the European Union’s Horizon 2020 research and innovation programme under the Marie Skłodowska-Curie grant agreement No 956071 AppQInfo - Applications And Hardware For Photonic Quantum Information Processing. 

R.~Yehia acknowledges funding from the European Union (ERC, ASC-Q, 101040624) and the Horizon Europe program (QUCATS).  It is also supported by the Government of Spain (Severo Ochoa CEX2019-000910-S and FUNQIP), Fundació Cellex, Fundació Mir-Puig, Generalitat de Catalunya (CERCA program).

M.~Schiavon acknowledges funding from the European Union's Horizon 2020 research and innovation programme under grant agreement No 101082596 (project QUDICE).

Views and opinions expressed are however those of the author(s) only and do not necessarily reflect those of the European Union or the European Research Council. Neither the European Union nor the granting authority can be held responsible for them.

\printbibliography

\newpage
\section*{Appendices}

\appendix
\counterwithin*{equation}{section}
\renewcommand\theequation{\thesection\arabic{equation}}

\section{Geometrical considerations}
\label{sec:geocons}

We give a brief description of the geometrical considerations relevant for the definition of the free-space models developed in this work. In particular, we present a general scenario based on downlink slant propagation as illustrated in Fig.~\ref{fig:prezfig}. The optical ground station is positioned at an altitude $h_0$ with respect to sea level, and the balloon at an altitude $H$ and zenith angle $\theta_z$ at the frame of reference of the ground station. Moreover, we assume that Earth is perfectly spherical with radius $R_E = 6371$ km. The propagation distance $z$ between the ground station and the balloon is given by 
\begin{equation}
z=\sqrt{R_{\mathrm{B}}^2+R_{\mathrm{G}}^2\left[\cos ^2 \left(\theta_z\right)-1\right]}-R_{\mathrm{G}} \cos\left(\theta_z\right),
\end{equation}
where $R_B = R_E + H$ and $R_G = R_E + h_0$. Alternatively, we may compute $z$ using arc lengths $s$ between points on the Earth's surface (for example distances between cities) according to 
\begin{equation}
z = \sqrt{R_{\mathrm{G}}^2+R_{\mathrm{S}}^2-2 R_{\mathrm{G}} R_{\mathrm{S}} \cos \left(\theta_s\right)},
\end{equation}
where $\theta_s = s/R_E$ is the subtending angle, as shown in Fig.~\ref{fig:prezfig}.

In the horizontal channel (Fig.~\ref{fig:horiz}), we consider both the height of the balloons $H$ and the minimal height of the path between the two balloons $h_{min}$. The minimal height can be calculated through
\begin{equation}
    h_{\rm min} = \mathrm{cos}(\theta_s)\left(H + R_E\right) - R_E,
\end{equation}
where the subtending angle can be found through the distance $z$ between the balloons as 
\begin{equation}
    \theta_s = \mathrm{arcsin}\left[\frac{z}{2(H+ R_E)}\right].
\end{equation}

Whenever possible, we exploit formulas extracted for Gaussian beams, whose complex amplitude is given by \cite{saleh2019fundamentals}
\begin{equation}
U(\mathbf{R})=A_0 \frac{W_0}{W(z)} \exp \left[-\frac{r^2}{W^2(z)}\right] \exp \left[-j k z-j k \frac{r^2}{2 F(z)}+j \zeta(z)\right],
\end{equation}
where z is the distance from the waist of the beam, $\mathbf{R}$ is the three-dimensional position vector, $r^2 = x^2 + y^2$, $W_0$ is the beam waist radius at the transmitter and $W(z)$ the waist radius after propagation of length $z$ that is broadened due to diffraction. The latter is given by
\begin{equation} \label{eq:freespace_beam}
    W(z) = W_0 \sqrt{1 + \left(\frac{\lambda z}{\pi W_0^2}\right)^2},
\end{equation}
where $k = 2\pi/\lambda$ is the wavenumber with $\lambda$ the wavelength of the beam. The radius of curvature $F(z)$ is given by
\begin{equation} \label{eq:curvature_r}
F(z)=z\left[1+\left(\frac{z_0}{z}\right)^2\right],
\end{equation}
and since it is infinite at the transmitter plane ($z = 0$), this also means that the studied beams are collimated. Finally, $\zeta(z)$ is a phase retardation factor
\begin{equation}
\zeta(z)=\tan ^{-1} \frac{z}{z_0},
\end{equation}
with $z_0$ the Rayleigh range, given as
\begin{equation}
    z_0 = \frac{\pi W_0^2}{\lambda}.
\end{equation} 
Otherwise, for ease of calculation, we consider the spherical wave approximation that holds when the propagation distance is larger than the Rayleigh range. The expression for the spherical wave is \cite{saleh2019fundamentals}
\begin{equation}
U(\mathbf{R}) = \frac{A_0}{z} \exp (-j k z) \exp \left[-j k \frac{r^2}{2 z}\right],
\end{equation}
where $A_0$ is a constant. 

\section{Collection efficiency} \label{sec:app_pdt}

\subsection{Atmospheric scintillation} \label{sec:collection_scint}

Scintillation can be quantified either by the log-amplitude (Rytov) variance $\sigma_R^2$ or the scintillation index $\sigma_I^2$, which is the normalized variance of the intensity fluctuations. 
When $\sigma_R^2$ is less than 1, the turbulence is deemed weak, whereas it is regarded as strong when $\sigma_R^2$ exceeds 1. The Rytov variance for a channel distance $z$ in the horizontal case is given by \cite{andrews2005laser} 
\begin{equation}
    \sigma_{\mathrm{R}}^{2(\rm hor)}=1.23 C_{\mathrm{n}}^2 k^{7 / 6} z^{11 / 6}.
\end{equation}
In the downlink, when a spherical wave is considered, $\sigma_R^2$ is \cite{andrews2005laser}
\begin{equation}
    \sigma_R^{2(\rm down)} = 2.25 k^{7 / 6} \sec^{11 / 6} (\theta_z)\int_{h_0}^{H} C_n^2(h)\left(h-h_0\right)^{5 / 6}\left(\frac{H - h}{H - h_0}\right)^{5 / 6} dh,
\end{equation}
with sec being the secant function, $\theta_z$ the zenith angle of the aerial platform, $h_0$ the altitude of the ground station and $H$ the altitude of the aerial platform.
Under the Rytov approximation or under the weak fluctuation regime, $\sigma_R^2$ and $\sigma_I^2$ can be approximated as equal. This approximation however leads to overestimation of the scintillation index as the irradiance fluctuations become stronger. In particular, when the receiving aperture is larger than the irradiance correlation width $\rho_c$, the receiver experiences what is called aperture averaging, which is manifested as a reduction in the measured level of scintillation. The correlation width can be quantified as \cite{fante1975electromagnetic} 
\begin{equation}
    \rho_c^{(\rm weak)} = \sqrt{\lambda z},
\end{equation}
when turbulence is weak, or as
\begin{equation}
    \rho_c^{(\rm strong)} = 0.36 \cdot \sigma_R^{-3/5} \sqrt{\lambda z},
\end{equation}
when turbulence is strong. If $D_{\rm Rx}$ is indeed larger than $\rho_c$, we may consider the aperture-averaged scintillation index for a spherical wave \cite{andrews2005laser}
\begin{equation}
    \sigma_I^2\left(D_{\rm Rx}\right)^{(\rm sp)}=\exp \left[\frac{0.49 \beta_0^2}{\left(1+0.18 d^2+0.56 \beta_0^{12 / 5}\right)^{7 / 6}}+\frac{0.51 \beta_0^2\left(1+0.69 \beta_0^{12 / 5}\right)^{-5 / 6}}{1+0.90 d^2+0.62 d^2 \beta_0^{12 / 5}}\right]-1 ,
\end{equation}
where $d$ is 
\begin{equation}
    d = \sqrt{\frac{k D_{\rm Rx}^2}{4 z}},
\end{equation}
with $D_{\rm Rx}$ the aperture of the receiving telescope and $\beta_0^2$ the Rytov variance for a spherical wave, given by
\begin{equation}
    \beta_0^2 = 0.4065 \sigma_R^2.
\end{equation}

\subsection{Beam broadening}
After traveling some path of length $z$, a light beam with initial beam waist $W_0$ will broaden due to diffraction and turbulence. The turbulence-induced beam broadening impacts the long-term size of a propagating Gaussian beam $W_{\rm LT}(z)$ as \cite{andrews2005laser}
\begin{equation} \label{eq:lt_beam}
    W_{\rm LT}(z) = W(z)\sqrt{1 + 1.63\sigma_R^{12/5}\frac{2z}{kW^2(z)}}.
\end{equation}
Substituting Eq.~\eqref{eq:freespace_beam} into Eq.~\eqref{eq:lt_beam}, we get 
\begin{equation}
    W_{\rm LT}(z) = W_0\sqrt{1 + \left(\frac{\lambda z}{\pi W_0^2}\right)^2\ + 1.63 \sigma_R^{12/5}\frac{2z}{kW_0^2}}.
\end{equation}
The above formula is quite general, since although extracted at the strong irradiance fluctuation regime, it can be used to approximate weak turbulence conditions too.

\subsection{Beam wandering} \label{sec:beam_wandering}

The turbulence-induced beam wandering variance for a Gaussian beam propagating in a horizontal channel, also valid in the strong irradiance fluctuation regime, is \cite{andrews2005laser}
\begin{equation} \label{eq:wander_horiz}
\left\langle r_c^2\right\rangle ^{(\rm gb, hor)} = 7.25 C_n^2 z^3 W_0^{-1 / 3} \int_0^1 \frac{\xi^2}{\left[\left(\Theta_0+\bar{\Theta}_0 \xi\right)^2+1.63 \sigma_R^{12 / 5} \Lambda_0(1-\xi)^{16 / 5}\right]^{1 / 6}} d\xi,
\end{equation}
where $\Theta_0$, $\bar{\Theta}_0$ and $\Lambda_0$ are parameters characterizing the transmitted Gaussian beam through
\begin{equation}
\Theta_0=1-\frac{z}{R_0}, \ \bar{\Theta}_0 = 1 - \Theta_0, \  \Lambda_0= \frac{z}{z_0} = \frac{2 z}{k W_0^2}.
\end{equation}
Parameter $\Theta_0$ (which is equal to 1 for collimated beams) quantifies the curvature of the beam, and $\Lambda_0$ is also known as the Fresnel length. Eq.~\eqref{eq:wander_horiz} can be adapted to the downlink propagation through 
\begin{equation} \label{eq:wander_down}
\left\langle r_c^2\right\rangle^{(\rm gb, down)} = 7.25 \mathrm{sec}^3(\theta_z) W_0^{-1 / 3} \int_{h_0}^H \frac{C_n^2(h)(h - h_0)^2}{\left[\left(\Theta_0+\bar{\Theta}_0 \frac{h - h_0}{H - h_0}\right)^2+1.63 \sigma_R^{12 / 5} \Lambda_0\left(\frac{H - h}{H - h_0}\right)^{16 / 5}\right]^{1 / 6}} dh.
\end{equation}

\subsection{Probability distribution of collection efficiency} \label{sec:app_pdt_deriv}

The log-negative Weibull distribution, valid in the weak turbulence regime, is given by
\begin{equation}\label{eq:weibull}
\begin{aligned}
P_{\rm weak}(\eta_{D_{\rm Rx}})= & \frac{R^2}{\sigma_{\rm wander}^2 \eta_{D_{\rm Rx}} l}\left(\ln \frac{\eta_0}{\eta_{D_{\rm Rx}}}\right)^{2 / l-1} \exp \left[-\frac{R^2}{2 \sigma_{\rm wander}^2}\left(\ln \frac{\eta_0}{\eta_{D_{\rm Rx}}}\right)^{(2 / l)}\right].
\end{aligned}
\end{equation}
In the above, $\eta_0$ is the maximal transmittance of a Gaussian beam with short-term beam-spot width $W_{\rm ST}$ at a circular aperture with radius $r_{\rm Rx}$, obtained as
\begin{equation}
    \eta_0=1-\exp \left[-2 \frac{r_{\rm Rx}^2}{W_{\mathrm{ST}}^2}\right],
\end{equation}
while $l$ and $R$ parameters given by
\begin{equation} \label{eq:lambda_par}
    l = 8 \frac{r_{\rm Rx}^2}{W_{\mathrm{ST}}^2}\frac{e^{-4\left(r_{\rm Rx}^2 / W_{\mathrm{ST}}^2\right)} \mathrm{I}_1\left(4 \frac{r_{\rm Rx}^2}{W_{\mathrm{ST}}^2}\right)}{1-\exp \left[-4 \frac{r_{\rm Rx}^2}{W_{\mathrm{ST}}^2}\right] \mathrm{I}_0\left(4 \frac{r_{\rm Rx}^2}{W_{\mathrm{ST}}^2}\right)} \left[\ln \left(2 \frac{\eta_0}{1-\exp \left[-4 \frac{r_{\rm Rx}^2}{W_{\mathrm{ST}}^2}\right] \mathrm{I}_0\left(4 \frac{r_{\rm Rx}^2}{W_{\mathrm{ST}}^2}\right)}\right)\right]^{-1},
\end{equation}
and
\begin{equation} \label{eq:R_par}
    R= r_{\rm Rx}\left\{\ln \left[2 \frac{\eta_0}{1-\exp \left[-4 \frac{r_{\rm Rx}^2}{W_{\mathrm{ST}}^2}\right] \mathrm{I}_0\left(4 \frac{r_{\rm Rx}^2}{W_{\mathrm{ST}}^2}\right) }\right]\right\}^{-1 / l}
\end{equation}
respectively. The function $I_n(x)$ in Eq.~\eqref{eq:lambda_par} is the modified Bessel function of $n^{\rm th}$ order.

As mentioned in Sec.~\ref{sec:coll_eff}, in the general case concerning stronger fluctuations as well,
we can extract the collection efficiency \acrshort{pdf} using the law of total probability, which enables us to separate the contributions of beam-spot distortion $P\left(\eta_{\rm D_{Rx}} \mid \boldsymbol{r}\right)$ and beam wandering $\rho\left(\boldsymbol{r}\right)$ effects as
\begin{equation} \label{eq:pdt_vec_app}
    P_{D_{\rm Rx}}(\eta_{\rm D_{Rx}})=\int_{\mathbb{R}^2} P\left(\eta_{\rm D_{Rx}} \mid \boldsymbol{r}\right) \rho\left(\boldsymbol{r}\right) d^2 \boldsymbol{r},
\end{equation}
where the distribution of the random transverse vector $\boldsymbol{r}$ is described by the two-dimensional Gaussian function
\begin{equation} \label{eq:weibull_vec}
    \rho\left(\boldsymbol{r}\right)=\frac{1}{2 \pi \sigma_{\rm wander}^2} \exp \left[-\frac{\boldsymbol{r}^2}{2 \sigma_{\rm wander}^2}\right].
\end{equation}

The effects of beam-spot distortions incorporated in $P\left(\eta \mid \boldsymbol{r}\right)$, that are more prominent in longer links and stronger fluctuation conditions, can be approximated by the truncated log-normal distribution
\begin{equation}
    \begin{aligned}
        P(\eta_{\rm D_{Rx}}|\boldsymbol{r}) & \approx P_{\rm strong}(\eta_{\rm D_{Rx}} ; \mu, \sigma) \\
        & = \begin{cases}\frac{1}{\mathcal{F}(1)} \frac{1}{\sqrt{2 \pi} \eta_{\rm D_{Rx}} \sigma} \exp \left[-\frac{(\ln \eta_{\rm D_{Rx}}+\mu)^2}{2 \sigma^2}\right] & \text { for } \eta_{\rm D_{Rx}} \in[0,1] \\
        0 & \text { otherwise, }\end{cases}
    \end{aligned}
\end{equation}
with $\mu$ and $\sigma$

\begin{equation} \label{eq:pdt_mu_sigma}
\begin{gathered}
\mu=\mu\left(\langle\eta_{\rm D_{Rx}}\rangle,\left\langle\eta_{\rm D_{Rx}}^2\right\rangle\right) \approx-\ln \left[\frac{\langle\eta_{\rm D_{Rx}}\rangle^2}{\sqrt{\left\langle\eta_{\rm D_{Rx}}^2\right\rangle}}\right], \\
\sigma=\sigma\left(\langle\eta_{\rm D_{Rx}}\rangle,\left\langle\eta_{\rm D_{Rx}}^2\right\rangle\right) \approx \sqrt{\ln \left[\frac{\left\langle\eta_{\rm D_{Rx}}^2\right\rangle}{\langle\eta_{\rm D_{Rx}}\rangle^2}\right]},
\end{gathered}
\end{equation}
and where $F(1)$ is the cumulative function of the non-truncated log-normal distribution at the point $\eta_{D_{\rm Rx}} = 1$. Combining Equations \eqref{eq:pdt_vec_app} and \eqref{eq:weibull_vec}, the PDF can also be expressed in polar coordinates as 
\begin{equation} \label{eq:pdt}
    P_{D_{\rm Rx}}(\eta_{D_{\rm Rx}}) = \frac{1}{\sigma_{\rm wander}^2}\int_0^{\infty} P_{\rm strong}(\eta_{D_{\rm Rx}} ; \mu, \sigma) \exp \left[-\frac{r^2}{2 \sigma_{\rm wander}^2}\right] r dr.
\end{equation} 
Under the assumption of weak beam wandering, which we implement in our software by assuming that $\sigma_{\rm wander}^2$ is smaller than the aperture radius of the receiving telescope, the average collection efficiency $\langle\eta_{D_{\rm Rx}}\rangle$ is
\begin{equation} \label{eq:pdt_eta_mean}
    \langle\eta_{D_{\rm Rx}}\rangle = \eta_0 \, \mathrm{exp} \left[-\left(\frac{r}{R}\right)^{\lambda}\right],
\end{equation}
and the second conditional moment can be calculated through the approximation 
\begin{equation} \label{eq:pdt_eta_2_mean}
    \langle\eta_{D_{\rm Rx}}^2\rangle = \langle\eta_{D_{\rm Rx}}\rangle^2 \left[1 + \sigma_I^2(D_{\rm Rx})\right].
\end{equation}

\section{Fiber-coupling efficiency} \label{sec:app_smf_pdf}

For the distances that we study in this work, the radius of curvature of spherical waves is approximately equivalent to the channel distance~\cite{saleh2019fundamentals}. Since the receiver aperture is generally much smaller than the considered propagation distances, we can assume that the optical beam entering the receiving system has the form of a plane wave. After propagation in the free-space channel, the complex optical field in the form of an unbounded plane wave is given, according to Rytov theory, by \cite{andrews2005laser} 
\begin{equation} \label{eq:unbounded_plane}
U(\mathbf{r}, t)=U_0(\mathbf{r}, t) \exp [\Psi(\mathbf{r}, t)],
\end{equation}
where $U_0(\mathbf{r}, t)$ is the unperturbed optical field and $\Psi(\mathbf{r}, t) = \Psi_1(\mathbf{r}, t)+ \Psi_2(\mathbf{r}, t) + \dots$ is the complex phase perturbation of the form $ \Psi_i(\mathbf{r}, t) = \chi_i(\mathbf{r}, t)+j \phi_i(\mathbf{r}, t)$. An alternative way to express Eq.~\eqref{eq:unbounded_plane} is through
\begin{equation}
U(\mathbf{r}, t)=U_0(\mathbf{r}, t) \exp [\chi(\mathbf{r}, t)+j \phi(\mathbf{r}, t))],
\end{equation}
from which it can be seen that $\chi(\mathbf{r}, t)$ encapsulates the log-amplitude fluctuations and $\phi(\mathbf{r}, t)$ the phase aberrations.

When the turbulence is weak, the complex optical field can be expressed in terms of the 1\textsuperscript{st} order Rytov approximation, which considers that $U$ is a circular Gaussian random variable, hence $\chi = \chi_1$ and $\phi = \phi_1$ are also Gaussian. The irradiance (intensity) of the field is then
\begin{equation}
    I=\left|U_0\right|^2 \exp \left(\Psi+\Psi^*\right)= A^2 e^{2 \chi},
\end{equation}
where $A$ the amplitude of the unperturbed field. It follows from this equation that the irradiance is lognormally distributed, since $\chi$ is a Gaussian random variable.

On the other hand, in the general case of weak-to-strong atmospheric turbulence, the 2\textsuperscript{nd} order Rytov approximation should be used for the calculations, so $\chi = \chi_1 + \chi_2$ and $\phi = \phi_1 + \phi_2$. The irradiance then takes the form
\begin{equation}
    I = A^2 e^{2\chi_1}e^{2\chi_2},
\end{equation}
in which the term $e^{2\chi_2}$ essentially performs a random modulation of $A^2e^{2\chi_1}$, and $I$ can be more accurately represented by a Gamma-Gamma distribution. However, under aperture averaging, it has been demonstrated \cite{vetelino2007fade} that the effect of $e^{2\chi_2}$ is negligible and the irradiance fluctuations can still be described as lognormal. 

The matching between the incident beam and the SMF mode at the pupil of the receiving telescope is given by the overlap integral \cite{canuet2018statistical}
\begin{equation} \label{eq:overlap_int}
    \Omega=\frac{\left\langle U \mid M_0\right\rangle_P}{\left[\langle U \mid U\rangle_P \times\left\langle M_0 \mid M_0\right\rangle_P\right]^{1 / 2}},
\end{equation}
where $M_0$ is the SMF mode expressed in the pupil plane, and $P$ is the pupil transmittance function
\begin{equation}
P(\mathbf{r})=\left\{\begin{array}{ll}
1 & \text { if } \alpha_{\rm obs} \leq \frac{2|r|}{D} \leq 1 \\
0 & \text { otherwise }
\end{array} .\right.
\end{equation}
with $\alpha_{\rm obs} = D_{\rm obs}/D_{\rm Rx}$ the obstruction ratio, where $D_{\rm obs}$ is the diameter of the obstruction in front of the receiver aperture. The operator $\langle\cdot \mid \cdot\rangle_Z$ in Eq.~\eqref{eq:overlap_int} refers to the scalar product
\begin{equation}
\langle X \mid Y\rangle_Z \triangleq \iint Z(\mathbf{r}) \cdot X(\mathbf{r}) \cdot Y^*(\mathbf{r}) \mathrm{d}^2 \mathbf{r}.
\end{equation}

The fiber-coupling efficiency $\eta_{\rm SMF}$ is the square modulus of the overlap integral
\begin{equation}
    \eta_{\rm SMF} = |\Omega|^2,
\end{equation}
and can be more conveniently calculated by the ratio 
\begin{equation} \label{eq:coupl_int_unwrapped}
    \frac{\eta_{\rm SMF}}{\eta_0} = \left|\frac{\Omega}{\Omega_0}\right|^2 = \left|\frac{\left\langle U \mid M_0\right\rangle_P}{\left\langle U_0 \mid M_0\right\rangle_P}\right|^2 = \left|\frac{\iint\mathrm{exp}[\chi(\mathbf{r}, t)+j \phi(\mathbf{r}, t))] M_0(\mathbf{r}) P(\mathbf{r}) \mathrm{d^2}\mathbf{r}}{\iint M_0(\mathbf{r}) P(\mathbf{r}) \mathrm{d^2}\mathbf{r}} \right|^2,
\end{equation}
where $\eta_0$ and $\Omega_0$ are the maximum achievable coupling efficiency and overlap integral without the presence of turbulence respectively. 

If we assume statistical independence of scintillation and phase effects due to the different origins of log-amplitude and phase fluctuations \cite{perlot2007turbulence}, we can study the contributions of $\chi(\mathbf{r}, t)$ and $\phi(\mathbf{r}, t)$ in Eq.~\eqref{eq:coupl_int_unwrapped} separately. Moreover, since the collection efficiency PDF already includes the decrease in collected power due to fluctuations in the wavefront intensity profile caused by scintillation, we approximate the impact on the coupling efficiency by considering an average value for the beam irradiance, which enables us to move $\mathrm{exp}[\chi(\mathbf{r}, t)]$ out of the integral. Hence, $\eta_{\rm SMF}$ consists of three terms
\begin{equation} \label{eq:eta_smf}
    \eta_{\rm SMF} = \eta_0 \eta_{\rm \chi} \eta_{\phi},
\end{equation}
where $\eta_{\rm \chi}$ is the term corresponding to the contribution of scintillation, and $\eta_{\phi}$ the coupling efficiency due to wavefront aberrations which can be partially corrected by AO. We will now look at each of these terms one by one.

\subsection{Coupling efficiency in the absence of turbulence}

Even in the diffraction-limited case, there is mismatch between the unperturbed beam arriving at the receiving telescope and the mode of the fiber (backpropagated to the aperture pupil). The resulting efficiency of the coupling $\eta_0$ depends on the obstruction ratio according to \cite{ruilier2001coupling}
\begin{equation}\label{eq:eta_0}
\eta_0(\alpha_{\rm obs}, \beta)=2\left[\frac{\exp \left(-\beta^2\right)-\exp \left(-\beta^2 \alpha_{\rm obs}^2\right)}{\beta\left(1-\alpha_{\rm obs}^2\right)^{1 / 2}}\right]^2.
\end{equation}
Parameter $\beta$ in Eq.~\eqref{eq:eta_0} is given by 
\begin{equation} \label{eq:beta_coupling}
\beta=\frac{\pi D_{\mathrm{Rx}}}{4 \lambda} \frac{\mathrm{MFD}}{f_{\rm SMF}},
\end{equation}
with $\mathrm{MFD}$ the mode field diameter of the fiber and $f_{\rm SMF}$ the focal length of the collimating lens in front of the \acrshort{smf}. It can be chosen appropriately to optimize the maximum fiber coupling efficiency $\eta_0$ and has a value of 1.12 in the case of no obstruction.

\subsection{Impact of atmospheric scintillation}

As already explained, we can assume that $\chi$ is a Gaussian random variable even in the moderate-to-strong fluctuation regime when there is aperture averaging. We approximate the impact of scintillation on the coupling efficiency through the average value of $\chi$ according to \cite{fried1967aperture}
\begin{equation} \label{eq:logamp_scint}
\eta_{\rm \chi} \approx \exp \left[-C_\chi(0)\right]=\exp \left[-\sigma_\chi^2\right],
\end{equation}
where $C_{\chi}(r)$ is the log-amplitude spatial covariance function, and $\sigma_{\chi}^2$ is the log-amplitude variance which in the general case, including moderate-to-strong fluctuations, is given by \cite{vetelino2007fade}
\begin{equation}
\sigma_{\chi}^{2(\rm down)}=\ln \left(\sigma_I^2+1\right).
\end{equation}
For the horizontal channels with weak turbulence that we study here, the formula for $\sigma_{\chi}^2$ is 
\begin{equation}
    \sigma_{\chi}^{2(\rm hor)}=\left[1 + \sigma_R^{2(\rm hor)}\right]^{-1/4},
\end{equation}
where it is implied that $\sigma_I^2 = \sigma_R^2$.

\subsection{Impact of wavefront aberration and adaptive optics} \label{sec:phase_term}

At this point, it is also convenient to define the instantaneous spatial average and variance operators as \cite{canuet2018statistical}
\begin{equation}
\langle X\rangle_Z \triangleq \frac{\langle X \mid 1\rangle_Z}{\langle 1 \mid 1\rangle_Z},
\end{equation}
and 
\begin{equation}
\sigma_Z^2(X) \triangleq\left\langle X^2\right\rangle_Z-\langle X\rangle_Z^2
\end{equation}
respectively. Defining $W(\mathbf{r}) = M_0(\mathbf{r}) P(\mathbf{r})$ as the weighted pupil function and focusing on the phase term, we have that
\begin{equation}
    \eta_\phi = \left|\frac{\Omega_{\phi}}{\Omega_0}\right|^2 = \left|\frac{\iint\mathrm{exp}[j \phi(\mathbf{r}, t))] M_0(\mathbf{r}) P(\mathbf{r}) \mathrm{d^2}\mathbf{r}}{\iint M_0(\mathbf{r}) P(\mathbf{r}) \mathrm{d^2}\mathbf{r}}\right|^2 = \left|\left\langle\mathrm{exp}(j \phi)\right\rangle_W\right|^2.
\end{equation}
The exponential term in the above equation can be rewritten as 
\begin{equation}
    \left|\left\langle\mathrm{exp}(j \phi)\right\rangle_W\right|^2 = \left|\mathrm{exp}(j\langle\phi\rangle_W) \langle\mathrm{exp}[j(\phi - \langle \phi \rangle_W)]\rangle_W\right|^2 = \left|\langle\mathrm{exp}[j(\phi - \langle \phi \rangle_W)]\rangle_W\right|^2,
\end{equation}
and then we expand the remaining term in its Taylor series 
\begin{equation}
    \left|\langle\mathrm{exp}[j(\phi - \langle \phi \rangle_W)]\rangle\right|^2 = \left|\left\langle 1 + j(\phi - \langle \phi \rangle_W) + \frac{[j(\phi - \langle \phi \rangle_W)]^2}{2} + \dots \right\rangle_W\right|^2.
\end{equation}
When $\phi - \langle \phi \rangle_W \ll 1$, we end up with 
\begin{equation}
     \eta_{\phi} \simeq \left|\mathrm{exp}\left[\frac{-\sigma_W^2(\phi)}{2}\right]\right|^2 = \mathrm{exp}[-\sigma_W^2(\phi)].
\end{equation}

\begin{figure}[h!]
    \centering
    \includegraphics[width=0.6\linewidth]{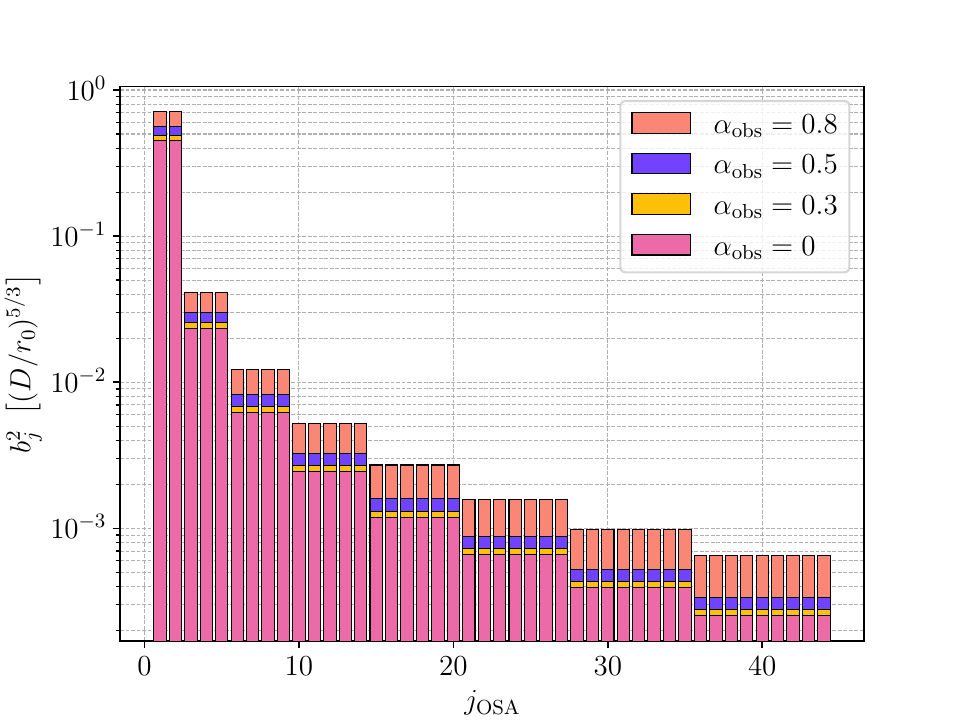}
    \caption{Annular Zernike coefficient variances in units of $\left(\frac{D}{r_0}\right)^{5/3}$ for different obstruction ratios.}
    \label{fig:bn2_annular}
\end{figure}

To derive an expression for variance $\sigma_W^2(\phi)$, we need to expand the aberrated phase $\phi(\mathbf{r})$ to a set of orthonormal polynomials. For circular pupils this is typically done using a Zernike polynomial basis \cite{noll1976zernike}. However, we want to consider the general case in which the pupil has a central obstruction, so we decompose the wavefront with annular Zernike polynomials instead. The decomposition can be written in normalised polar coordinates as \cite{dai2006zernike}
\begin{equation}
\phi\left(\frac{2r}{D_{\rm Rx}}, \theta, t ; \alpha_{\rm obs}\right)=\sum_{\substack{n,m \\ n=0}}^{\infty} b_n^m(t;\alpha_{\rm obs}) Z_n^m \left(\frac{2r}{D_{\rm Rx}}, \theta ; \alpha_{\rm obs}\right),
\end{equation}
where  $Z_n^m$ is the annular Zernike polynomial of radial degree $n$ and azimuthal degree $m$ and $b_n^m$ the coefficient associated with the $Z_n^m$ polynomial. The coefficients can also be labelled according to the single-index OSA/ANSI convention \cite{thibos2002standards} through 
\begin{equation}
    j=\frac{n(n+2)+m}{2}.
\end{equation}
The variances of the coefficients are given by \cite{dai2006zernike}
\begin{equation} \label{eq:annular_bj2}
\begin{aligned}
    \left\langle b_j^2\right\rangle = \left\langle\left|b_m^n(t;\alpha_{\rm obs})\right|^2\right\rangle &= \frac{0.023(n+1) \Gamma(n-5 / 6) \pi^{8 / 3}}{2^{5 / 3} \Gamma(17 / 6)\left(1-\alpha_{\rm obs}^2\right)\left[1-\alpha_{\rm obs}^{2(n+1)}\right]}\left(\frac{D_{\rm Rx}}{r_0}\right)^{5 / 3} \\
    & \times\left[\frac{\left(1+\alpha_{\rm obs}^{2 n+17 / 3}\right) \Gamma(14 / 3)}{\Gamma(17 / 6) \Gamma(n+23 / 6)}-\frac{2 \alpha_{\rm obs}^{2(n+1)}}{(n+1)!}\right. \\
    &\left.\times F_1\left(n-\frac{5}{6},-\frac{11}{6} ; n+2 ; \alpha_{\rm obs}^2\right)\right],
\end{aligned}
\end{equation}
where $\Gamma$ is the Gamma function, $F_1$ is the Gaussian hypergeometric function and $r_0$ the coherence width, also known as Fried's parameter \cite{fried1966optical}. The atmospheric coherence width quantifies the area of phase coherence of a distorted wavefront, or the maximum aperture size of a telescope to have diffraction-limited resolution, and is widely used to characterize the performance of imaging systems. The coherence width for a gaussian wave in the downlink case is given by \cite{andrews2005laser}
\begin{equation}
r_0^{(\rm g b, down)}=\left[\frac{\cos \left(\theta_z\right)}{0.423 k^2\left(\mu_{1 d}+0.622 \mu_{2 d} \Lambda^{11 / 6}\right)}\right]^{3 / 5},
\end{equation}
where 
\begin{equation}
\mu_{1 d}=\int_{h_0}^H C_n^2(h)\left[\Theta+\bar{\Theta}\left(1-\frac{h-h_0}{H-h_0}\right)\right]^{5 / 3} d h,
\end{equation}
\begin{equation}
\mu_{2 d}=\int_{h_0}^H C_n^2(h)\left(\frac{h-h_0}{H-h_0}\right)^{5 / 3} d h,
\end{equation}
and parameters $\Theta$, $\bar{\Theta}$ and $\Lambda$ characterize the output plane of a Gaussian beam according to
\begin{equation}
\Theta=\frac{1}{1 + \Lambda_0^2}, \ \bar{\Theta} = 1 - \Theta, \  \Lambda= \frac{\Lambda_0}{1 + \Lambda_0^2}
\end{equation}
when the beam is collimated. For the horizontal case and under weak irradiance fluctuations, the expression is
\begin{equation}
r_0^{(\rm g b, hor)}=\left[\frac{8}{3\left(a+0.618 \Lambda^{11 / 6}\right)}\right]^{3 / 5} \cdot\left[0.423 \cdot C_n^2\left(\frac{2 \pi}{\lambda}\right)^2 z\right]^{-3 / 5},
\end{equation}
where a is a parameter depending on the Gaussian beam curvature as 
\begin{equation}
a= \begin{cases}\frac{1-\Theta^{8 / 3}}{1-\Theta} & \Theta \geq 0 \\ \frac{1+\Theta^{8 / 3}}{1-\Theta} & \Theta<0\end{cases}
\end{equation}
The change of the normalized annular Zernike coefficient variances with the size of the central obstruction is illustrated in Fig.~\ref{fig:bn2_annular}.

It should be noted that this basis might be orthonormal for an annular pupil, but not for the weighted pupil function which includes both the central obstruction and the SMF mode. With the help of a conversion matrix $\mathbf{M}$, the annular basis coefficients can be transformed to a new set of coefficients that are orthonormal to the weighted pupil according to \cite{dai2007nonrecursive,dai2008orthonormal}
\begin{equation}
\mathbf{a}=\left(\mathbf{M}^T\right)^{-1} \mathbf{b}.
\end{equation}
The process for calculating the conversion matrix can be found in \cite{canuet2018statistical}. This matrix is almost diagonal, hence we will not take it into account as a first approximation.

Finally, the wavefront variance $\sigma_W^2(\phi)$ is expressed as a sum of the variances of the coefficients $\left\langle|b_n^m|^2\right\rangle$ and $\eta_{\phi}$ is
\begin{equation}
    \eta_{\phi} = \mathrm{exp}\left(-\sum_{\substack{n,m \\ n=1}}^{N_{\rm max}}  \left\langle|b_n^m|^2\right\rangle\right),
\end{equation}
where $N_{\rm max}$ is the maximum radial order used in the decomposition, that needs to be chosen large enough to minimize as much as possible the turbulent wavefront aberrations that are not taken into account. 

\begin{figure}[h!]
    \centering
    \includegraphics[width=0.6\linewidth]{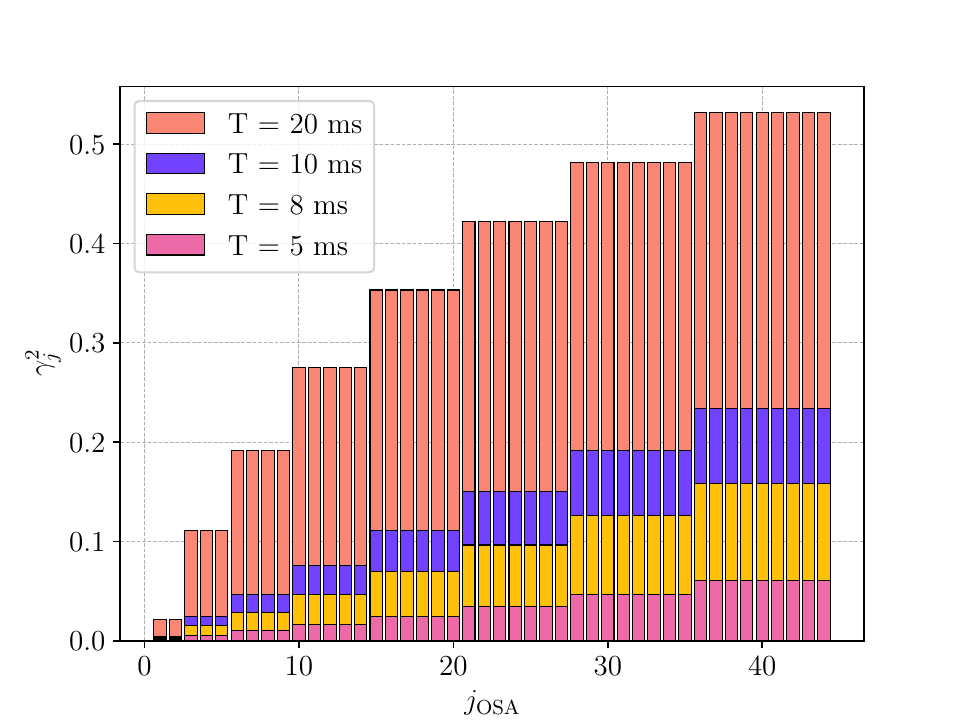}
    \caption{Attenuation coefficients for a system with $D_{\rm Rx} = 0.4$ m, $K_I = 1$, $\tau = 2T$ and channel wind-speed $\bar{v} = 10$ m/s, for different integration times $T$.}
    \label{fig:gamma_j2}
\end{figure}

The presence of an adaptive optics system impacts polynomials up to the radial order $N_{\rm AO}$. When the correction is perfect, we have that $\eta_{\phi}$ depends on the residual phase
\begin{equation}
    \eta_{\phi} = \mathrm{exp}\left(-\sum_{\substack{n,m \\ n = N_{\rm AO} + 1}}^{N_{\rm max}} \left\langle|b_n^m|^2\right\rangle\right).
\end{equation}
In practice, AO systems have finite bandwidth and cannot fully suppress the terms up to $N_{\rm AO}$. For that reason, we follow the treatment in \cite{scriminich2022optimal} (originally in \cite{roddier1999adaptive}) by introducing attenuation factors as
\begin{equation} \label{eq:attenuation_coeffs}
    \eta_{\phi} = \mathrm{exp}\left(-\sum_{\substack{n,m \\ n=1}}^{N_{\rm max}}  \gamma_n^2 \left\langle|b_n^m|^2\right\rangle\right),
\end{equation}
where $\gamma_n^2 = 1$ when $n > N_{\rm AO}$ and 
\begin{equation}
\gamma_n^2=\frac{\int\left|W_n(\tilde{\nu})\right|^2|\varepsilon(\tilde{\nu})|^2 \mathrm{~d} \tilde{\nu}}{\int\left|W_n(\tilde{\nu})\right|^2 \mathrm{~d} \tilde{\nu}},
\end{equation}
when $\gamma_n^2 \leq N_{\rm AO}$. Term $\left|W_n(\tilde{\nu})\right|^2$ in Eq.~\eqref{eq:attenuation_coeffs} is the power spectral density (PSD) of the temporal spectrum of n\textsuperscript{th} order aberrations, $\tilde{\nu}$ is the frequency of the AO loop and $\epsilon(\tilde{\nu})$ is the transfer function between the residual phase and aberrated wavefront fluctuations, which depends on the open-loop transfer function $G(\tilde{\nu})$ through
\begin{equation}
\varepsilon(\tilde{\nu})=\frac{1}{1+G(\tilde{\nu})}.
\end{equation}
A typical AO system can be modelled as a pure integrator with a Shack-Hartmann wave-front sensor (WFS) and a deformable mirror for correction. The open-loop transfer function is then
\begin{equation}
G(\tilde{\nu})=K_I \frac{\mathrm{e}^{-\tau \tilde{\nu}}\left(1-\mathrm{e}^{-T \tilde{\nu}}\right)}{(T \tilde{\nu})^2},
\end{equation}
where $K_I$ is the integral gain, $T$ is the WFS integration time and $\tau$ the latency of the control-actuator stage, that we consider to be $\tau = 2T$. The temporal power spectra are in turn given by \cite{conan1995wave}
\begin{equation}
\left|W_n(\tilde{\nu})\right|^2 \sim \begin{cases}\tilde{\nu}^{-2 / 3} & \tilde{\nu} \leqslant \tilde{\nu}_{\mathrm{c}}^{(n)}, n=1 \\ \tilde{\nu}^0 & \tilde{\nu} \leqslant \tilde{\nu}_{\mathrm{c}}^{(n)}, n \neq 1 \\ \tilde{\nu}^{-17 / 3} & \tilde{\nu}>\tilde{\nu}_{\mathrm{c}}^{(n)}\end{cases},
\end{equation}
where $\tilde{\nu}_{\mathrm{c}}^{(n)}$ is a cut-off frequency given by
\begin{equation}
\tilde{\nu}_{\mathrm{c}}^{(n)}=\frac{0.3(n+1) \bar{v}_{\rm wind}}{D_{\mathrm{Rx}}}.
\end{equation}
An example of how the attenuation coefficients depend on the finite bandwidth of the AO control loop is demonstrated in Fig.~\ref{fig:gamma_j2}.

\subsection{Average fiber coupling efficiency}
\label{sec:AvrgFiberCoupl}
In order to have an estimate of the average fiber-coupling efficiency based on
\begin{equation} \label{eq:phase_coupl_theor_mean}
    \langle\eta_{\rm SMF}\rangle = \left\langle\eta_0 \eta_{\chi}\eta_{\phi}\right\rangle = \eta_0 \eta_{\chi} \langle \eta_{\phi} \rangle,
\end{equation}
we can use formula \cite{scriminich2022optimal}
\begin{equation} 
\left\langle\eta_{\phi}\right\rangle=\prod_{\substack{n, m \\ n \leqslant N_{\rm AO}}} \frac{1}{\sqrt{1+2 \gamma_n^2\left\langle b_n^{m 2}\right\rangle}}+\prod_{\substack{n, m \\ n>N_{\rm AO}}} \frac{1}{\sqrt{1+2\left\langle b_n^{m 2}\right\rangle}},
\end{equation}
for the term attributed to wavefront perturbations and the impact of AO correction. We need to be aware, however, that the fiber-coupling efficiency model is accurate when the condition $\phi - \langle \phi \rangle_W \ll 1$ holds (Sec.~\ref{sec:phase_term}), or equivalently when $\sigma_W^2(\phi)$ is sufficiently small. As the total residual wavefront variance after correction increases, the error between the mean value calculated using Eq.~\eqref{eq:phase_coupl_theor_mean} and the one calculated directly from the PDF of Eq.~\eqref{eq:smf_pdf} becomes larger. If we express the standard deviation of the Zernike coefficients in terms of wavelength through
\begin{equation}
    \sigma_j^{\lambda-unit}=\frac{\sqrt{\left\langle b_j^2\right\rangle}}{2\pi},
\end{equation}
 and using the Rayleigh criterion \cite{malacara2007optical}, we can quantify a threshold above which the residual wavefront error is too high. According to this criterion, if $\sigma_j^{\lambda-unit}$ is smaller than $0.05\lambda$, then the performance of the aberrated system is virtually indistinguishable from that of an ideal, non-aberrated one. Indeed, if the maximum of the residual phase coefficients $\sigma_{j}^{\lambda-unit}$ is below the Rayleigh criterion, the error in the mean value is less than 10\%.

\section{Reciprocal modelling of uplink channel} \label{sec:uplink} 

Here we explain the modelling of the ground-to-balloon communication channel or uplink channel. For modelling the uplink channel, we base our analysis on the principle of reciprocity, which is typically used for uplink satellite communication. According to this principle, we can model the uplink channel using the same model as for the downlink, considering that the AO system at the ground station is pre-compensating for the aberrations during propagation. 

One way of pre-compensation in uplink channels is to have the receiver (the aerial platform) transmit a downlink beacon that can be used as a reference for the AO control loop. Anisoplanatism stems from the fact that, when a beam propagates through a specific portion of the atmosphere while the platform moves, it is distorted in a different manner compared to propagating in a different portion. Angular anisoplanatism starts to become more evident for angular separations larger than the isoplanatic angle $\theta_0$, which, for a Gaussian beam is given by \cite{andrews2005laser}:
\begin{equation}
\theta_0^{(\rm gb,up)}=\frac{\cos ^{8 / 5}(\theta_z)}{\left(H-h_0\right)\left[2.91 k^2\left(\mu_{1 u}+0.62 \mu_{2 u} \Lambda^{11 / 6}\right)\right]^{3 / 5}},
\end{equation}
where
\begin{equation}
\mu_{1 u}=\int_{h_0}^H C_n^2(h)\left[\Theta+\bar{\Theta}\left(\frac{h-h_0}{H-h_0}\right)\right]^{5 / 3} d h,
\end{equation}
\begin{equation}
\mu_{2 u}=\int_{h_0}^H C_n^2(h)\left(1-\frac{h-h_0}{H-h_0}\right)^{5 / 3} d h.
\end{equation}

If the angle between the downlink beacon and the uplink beam is $\theta$, the total wavefront error due to anisoplanatism can be approximated by \cite{fried1982anisoplanatism}:
\begin{equation}
    \sigma_{\rm aniso}^2 = \left(\frac{\theta}{\theta_0}\right)^{5/3}.
\end{equation}
In satellite communications for example, a ground-based transmitter needs to account for the satellite's movement during the time it takes for the tracking beam to travel back and forth between the satellite and the ground. The angle between the reference downlink beam and the pre-compensated uplink beam, called point-ahead, is then given by \cite{andrews2005laser}:
\begin{equation}
    \theta_{\rm pa} = \frac{2V_{\rm sat}}{c},
\end{equation}
where $V_{\rm sat}$ is the speed of the satellite, typically of the order of thousands of meters per second, and $c$ the speed of light. Differently from a satellite, a high-altitude balloon does not move at a constant high speed, but with a much lower speed of a few m/s and in random directions. If we assume that this random movement is Gaussian, centered around the position of the balloon when it sends the beacon and with a variance of a few meters, such anisoplanatism is only a very small fraction of a $\mu$rad, which is negligible.

\begin{figure}[ht!]
    \centering
    \includegraphics[width=0.4\linewidth]{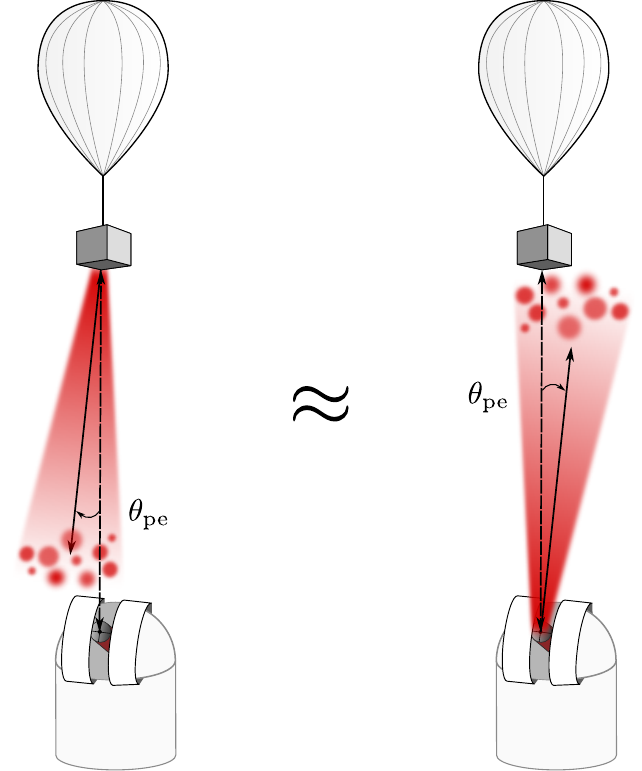}
    \caption{Reciprocal modelling of uplink communication between a ground station and a balloon. The pointing error in the downlink reference beacon is treated as pointing error stemming from the the ground station. This pointing error is used to quantify losses induced by anisoplanatism.}
    \label{fig:uplink_geometry}
\end{figure}

In this case, we can consider that the main source of anisoplanatism is the pointing error of the downlink reference beam, as illustrated in Fig.~\ref{fig:uplink_geometry}. If we assume that the downlink beacon channel has a pointing error $\theta_{\rm pe}$, as described in Sec.~\ref{sec:coll_eff}, and that there is no additional pointing error induced by the ground station in the uplink direction, we have a situation that is equivalent to a mispointing from the ground station transmitter of $\theta_{\rm pe}$. The anisoplanatic error becomes
\begin{equation}
    \sigma^2_{\rm aniso} = \left(\frac{\theta_{\rm pe}}{\theta_0}\right)^{5/3}.
\end{equation}

The impact of anisoplanatism in the uplink channel is finally introduced as an additional term in the total phase variance $\sigma_W^2(\phi)$, which is manifested as a fixed loss in the fiber-coupling efficiency according to \cite{lognoneOptimizationHighData2023}
\begin{equation}
    \eta_{\rm SMF} = \eta_0 \eta_{\rm \chi} \eta_{\rm \phi} \eta_{\rm aniso},
\end{equation}
where $\eta_{\rm aniso} = \mathrm{exp}\left(-\sigma^2_{\rm aniso}\right)$. This implies that, since the fiber coupling efficiency model we are using was originally developed for a downlink channel \cite{canuet2018statistical} and is accurate only for small values of the total phase variance, the requirements for the performance of the AO system are more stringent in the uplink case.

\section{Verification of the assumptions of the model for the simulations}
\label{sec:verifassumption}

As explained in Sec.~\ref{sec:setup}, there are a few conditions imposed by our model to ensure the accuracy of the simulation results. In the particular cases of our simulated networks, presented in Fig.~\ref{fig:trustedscenar} and \ref{fig:untrustedscenar} while considering the complete set of parameters in Table~\ref{tab:baselineparameters} and \ref{tab:QKDparameters}, we can check that the conditions are fulfilled. In Table~\ref{tab:conditions}, we show explicitly that these conditions are met for the simulated Italian architecture.
\begin{table}[!ht]
    \centering

\begin{tabular}{l|c|l}
    Name & Condition & Current value  \\ \hline
    Aperture averaging & $D_{\textrm{Rx}}>0.39130$ & $D_{\textrm{Rx}}=0.4$\\
    Rayleigh criterion & $\sigma_{j,\ \max}^{\lambda-unit}<0.05$ & $N_{\rm AO}=6 \rightarrow \sigma_{j,\ \max}^{\lambda-unit}=0.0368$ \\ 
    Small wandering & $\sigma_{\rm wander}<R_{\rm Rx}$  & $\sigma_{\rm wander}=0.242 \cdot R_{\rm Rx}$  \\
\end{tabular}
    \caption{Verification of the conditions of our free-space model.}
    \label{tab:conditions}
\end{table}
\end{document}